\documentclass{aa}  

\usepackage{graphicx}
\usepackage{natbib}
\usepackage{multirow,bigdelim} 
\usepackage{booktabs}  
\usepackage[colorlinks=true,linkcolor=blue,citecolor=blue]{hyperref}
\usepackage{txfonts}

\usepackage[separate-uncertainty, angle-symbol-over-decimal]{siunitx}

\graphicspath{{./Pictures/}} 

\usepackage[dvipsnames]{xcolor} 

\usepackage{longtable}
\usepackage{lscape}

\usepackage{orcidlink}

\let\orgautoref\autoref
\renewcommand{\autoref}
        {\def\equationautorefname{Eq.}%
         \def\figureautorefname{Fig.}%
         \def\sectionautorefname{Sect.}%
         \def\subsectionautorefname{Sect.}%
         \def\subsubsectionautorefname{Sect.}%
         \orgautoref}

\newcommand{\Pbday}{36-day}
\newcommand{\Pcday}{113-day}

\newcommand{\Pc}{113\,d}
\newcommand{\Pbfit}{$36.116\pm0.029\,\text{d}$}
\newcommand{\Pcfit}{$113.46\pm0.20\,\text{d}$}
\newcommand{\Mbfit}{$5.46\pm0.75\,\text{M}_\oplus$}
\newcommand{\Mcfit}{$9.7\pm1.9\,\text{M}_\oplus$}
\newcommand{\hoststar}{HN\,Lib}
\newcommand{\planetb}{HN\,Lib\,b}
\newcommand{\planetc}{HN\,Lib\,[c]}
\newcommand{\Prot}{$96\pm2$\,d}

\newcommand{\Protaliasone}{$\sim$125\,d}
\newcommand{\Protaliastwo}{$\sim$94\,d}

\newcommand*\samethanks[1][\value{footnote}]{\footnotemark[#1]}

\makeatletter
\renewcommand*\aa@pageof{, page \thepage{} of \pageref*{LastPage}}
\makeatother

\begin{document}

   \title{The CARMENES search for exoplanets around M dwarfs}

   \subtitle{A sub-Neptunian mass planet in the habitable zone of \hoststar}

   \author{E.~Gonz\'alez-\'Alvarez\,\orcidlink{0000-0002-4820-2053} \inst{\ref{inst:CSIC-INTA1}, \ref{inst:ucm}}\thanks{Both authors contributed equally to the manuscript.}
   \and J.~Kemmer \inst{\ref{inst:lsw}\samethanks{}}
   \and P.~Chaturvedi \inst{\ref{inst:TLS}}
   \and J.\,A.~Caballero\inst{\ref{inst:CSIC-INTA1}}   
   \and  {A.~Quirrenbach\inst{\ref{inst:lsw}}}   
   \and P.\,J.~Amado\inst{\ref{inst:IAA}}
   \and V.\,J.\,S.~B\'ejar\inst{\ref{inst:IAC},\ref{inst:ULL}}
   \and  {C.~Cifuentes\inst{\ref{inst:CSIC-INTA1}}}
   \and E.~Herrero \inst{\ref{inst:IEEC}}
   \and D.~Kossakowski \inst{\ref{inst:mpia}}
   \and A.~Reiners \inst{\ref{inst:IAG_goett}}
   \and I.~Ribas \inst{\ref{inst:IEEC-CSIC},\ref{inst:IEEC}}
   \and E.~Rodr\'iguez \inst{\ref{inst:IAA}}
   \and C.~Rodr\'iguez-L\'opez \inst{\ref{inst:IAA}}
   \and J.~Sanz-Forcada\inst{\ref{inst:CSIC-INTA1}}
   \and  {Y.~Shan\inst{\ref{inst:oslo},\ref{inst:IAG_goett}}}
   \and S.~Stock \inst{\ref{inst:lsw}}
   \and H.\,M.~Tabernero\inst{\ref{inst:CSIC-INTA1}}
   \and L.~Tal-Or \inst{\ref{inst:TelAviv},\ref{inst:IAG_goett}}
   \and M.\,R.~Zapatero~Osorio\inst{\ref{inst:CSIC-INTA1}}
   \and A.\,P.~Hatzes \inst{\ref{inst:TLS}}
   \and  {Th.~Henning\inst{\ref{inst:mpia}}}
   \and  {M.\,J.~L\'opez-Gonz\'alez\inst{\ref{inst:IAA}}}
   \and  {D.~Montes \inst{\ref{inst:ucm}}}
   \and J.\,C.~Morales\inst{\ref{inst:IEEC-CSIC},\ref{inst:IEEC}}
   \and E.~Pall\'e \inst{\ref{inst:IAC},\ref{inst:ULL}}
   \and  {S.~Pedraz \inst{\ref{inst:caha}}} 
   \and M.~Perger \inst{\ref{inst:IEEC-CSIC},\ref{inst:IEEC}}
   \and S.~Reffert \inst{\ref{inst:lsw}}
   \and  {S.~Sabotta \inst{\ref{inst:lsw}}}
   \and  {A.~Schweitzer \inst{\ref{inst:hamburg}}}
   \and  {M.~Zechmeister \inst{\ref{inst:IAG_goett}}}
}

        \institute{Centro de Astrobiolog\'ia, CSIC-INTA, Carretera de Ajalvir km 4, 28850 Torrej\'on de Ardoz, Madrid, Spain \label{inst:CSIC-INTA1}
        \and Departamento de F\'isica de la Tierra y Astrof\'isica \& IPARCOS-UCM (Instituto de F\'isica de Part\'iculas y del Cosmos de la UCM), Facultad de Ciencias F\'isicas, Universidad Complutense de Madrid, 28040 Madrid, Spain\label{inst:ucm}
        \and Landessternwarte, Zentrum f\"ur Astronomie der Universit\"at Heidelberg, K\"onigstuhl 12, 69117 Heidelberg, Germany \label{inst:lsw}
        \and Th\"uringer Landessternwarte Tautenburg, Sternwarte 5, D-07778 Tautenburg, Germany\label{inst:TLS}
        \and Instituto de Astrof\'isica de Andaluc\'ia (IAA-CSIC), Glorieta de la Astronom\'ia s/n, 18008 Granada, Spain \label{inst:IAA}
    \and Instituto de Astrof\'isica de Canarias, Avenida V\'ia L\'actea s/n, E-38205 La Laguna, Tenerife, Spain\label{inst:IAC}
    \and Departamento de Astrof\'isica, Universidad de La Laguna, E-38206 La Laguna, Tenerife, Spain\label{inst:ULL}  
    \and Institut de Ci\`encies de l'Espai (IEEC-CSIC), Campus UAB, Carrer de Can Magrans s/n, 08193, Bellaterra, Spain \label{inst:IEEC-CSIC} 
    \and Institut d'Estudis Espacials de Catalunya (IEEC), E-08034 Barcelona, Spain  \label{inst:IEEC}
    \and Max-Planck-Institut f\"ur Astronomie, K\"onigstuhl 17, 69117 Heidelberg, Germany\label{inst:mpia}
    \and Institut f\"ur Astrophysik, Georg-August-Universit\"at, Friedrich-Hund-Platz 1, 37077 G\"ottingen, Germany\label{inst:IAG_goett}
    \and Centre for Earth Evolution and Dynamics, Department of Geosciences, Universitetet i Oslo, Sem S{\ae}lands vei 2b, 0315 Oslo, Norway\label{inst:oslo} 
    \and Department of Physics, Ariel University, Ariel, 40700, Israel \label{inst:TelAviv}
    \and Centro Astron\'omico Hispano en Andalucia, Observatorio de Calar Alto, Sierra de los Filabres, 04550 G\'ergal, Spain\label{inst:caha}
    \and Hamburger Sternwarte, Gojenbergsweg 112, 21029 Hamburg, Germany\label{inst:hamburg}
    }

   \offprints{Esther~Gonz\'alez-\'Alvarez and Jonas Kemmer \\ \email{estgon11@ucm.es, jkemmer@lsw.uni-heidelberg.de}}
   \date{Received 28 February 2023 / Accepted 11 May 2023}

 
  \abstract
  {We report the discovery of \planetb{}, a sub-Neptunian mass planet orbiting the nearby ($d \approx$ = 6.25\,pc) M4.0\,V star \hoststar{} detected by our CARMENES radial-velocity (RV) survey. We {determined} a planetary minimum mass of $M_\text{b}\sin i = $ \Mbfit{} and an orbital period of $P_\text{b} = $ \Pbfit{}, using {$\sim$5\,yr of} CARMENES data, as well as archival RVs from HARPS and HIRES {spanning more than }13 years. The {flux received by the planet equals half the instellation on Earth}, which places it {in} the middle of the conservative habitable zone (HZ) of its host star. The {RV} data show evidence for another {planet candidate} with $M_\text{[c]}\sin i = $ \Mcfit{} and $P_\text{[c]} = $ \Pcfit{}. The long-term stability of the signal and the fact that the best model for our data is a two-planet model with an independent activity component stand as strong arguments for establishing a planetary origin. However, we cannot rule out stellar activity due to its proximity to  the rotation period of \hoststar{}, which we measured using CARMENES activity indicators and photometric data from a ground-based multi-site campaign as well as archival data. The discovery adds \planetb{} to the shortlist of super-Earth planets in the habitable zone of M dwarfs, but \planetc{} probably cannot be inhabited because, if confirmed, it would most likely be an icy giant.}


\maketitle
%

\section{Introduction}
\label{Introduction}

In the last decade, spectrographs delivering high-precision radial velocity (RV) measurements have reached the necessary precision to detect small planets close to (and even within) the habitable zone {\citep[HZ; e.g.,][]{Kasting1993, Kopparapu2014}} of late-type main-sequence stars. The {predicted} high number of super-Earth and Earth-like planets in the HZ {that could be detected} around M-dwarf {stars makes these stars} extremely interesting objects for planetary RV searches \citep{Udry2007,Zechmeister2009,Dressing2013,Bonfils2013,Tuomi2014,Sabotta2021}. Despite the significant number of detected planets around M-dwarf primaries, we are{} still far from fully understanding fundamental questions, such as how {these} planetary systems form and evolve {\citep{Ida2005,Raymond2007,Burn2021, Schlecker2022}.
Especially for the planets in the HZ, the formation history is important. This is because not only is the incoming stellar flux a crucial component in determining the actual habitability, but so are the properties of the planet itself, such as the composition of a potential atmosphere \citep{Tarter2007,Kaltenegger2017}. In this context, the increasing number of M-dwarf stars that harbor planets in the HZ (e.g., \citealt{SuarezMascareno2023} and \citealt{Kossakowski2023},  to cite the two most recent examples based on the radial velocity method) plays a significant role in exploring theories of planet formation and evolution.}

\begin{figure*}
\centering
\includegraphics{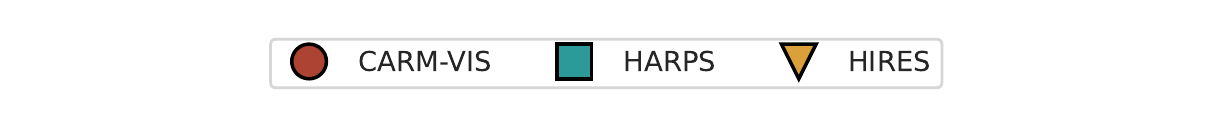}
\includegraphics{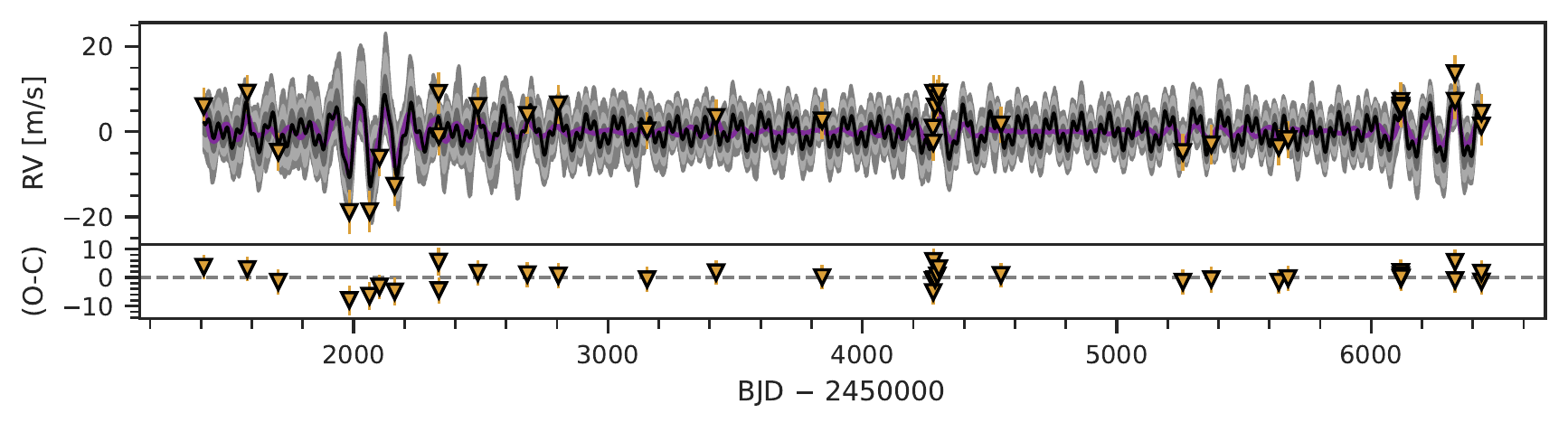}\\
\includegraphics{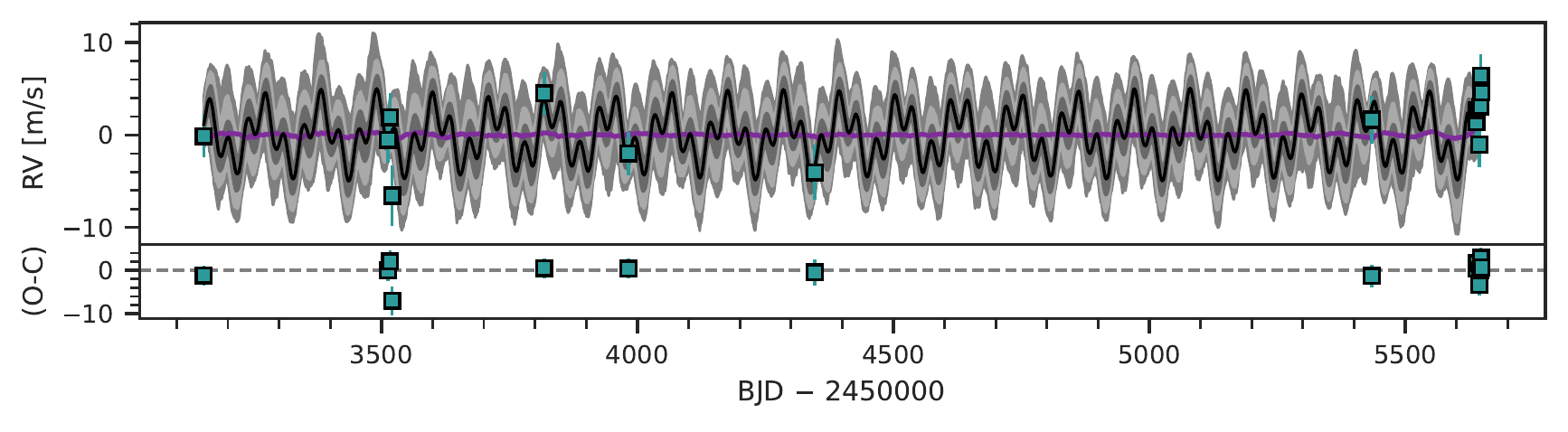}\\
\includegraphics{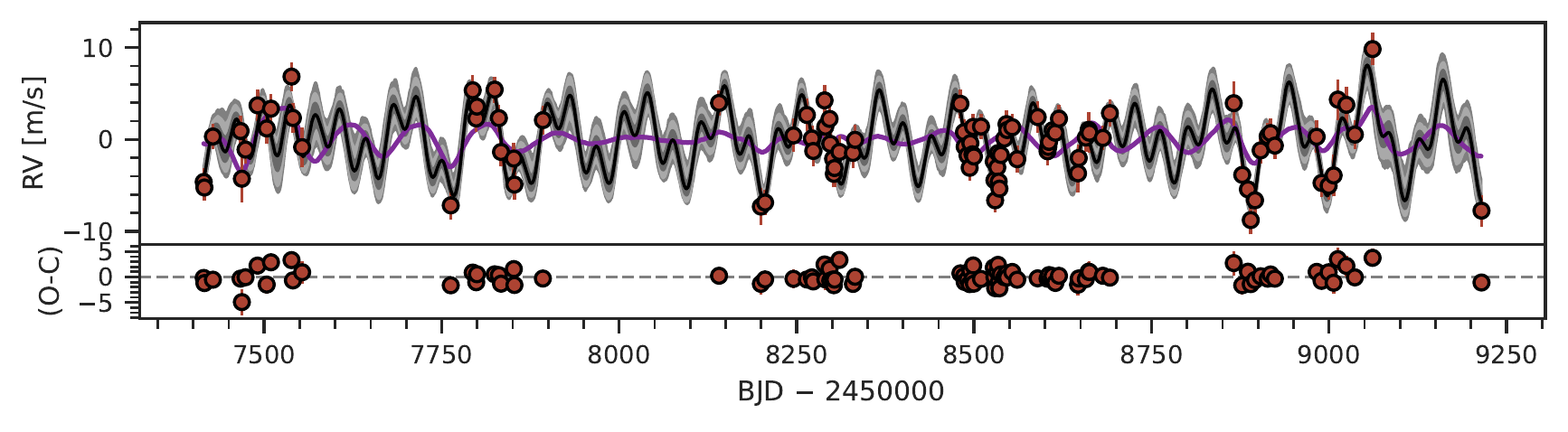}
\caption{RVs of \hoststar{} over time. The red, green, and yellow markers depict the CARMENES VIS, HARPS, and HIRES RV data, respectively.
The black lines show the median of \num{10000} samples from the posterior of the final {2P$_\text{(36\,d-ecc, 113\,d-circ)}$ + dSHO-GP$_\text{(96\,d)}$} model. Grey shaded areas denote the \SI{68}{\percent}, \SI{95}{\percent,} and \SI{99}{\percent} {confidence} intervals, respectively. The purple line shows the GP part of the model.
The instrumental RV offsets were subtracted from the measurements and the models. Error bars of the measurements include the jitter added in quadrature. The residuals after subtracting the median models are shown in the lower panels of each plot.} 
\label{fig:fit_rv_full}
\end{figure*}

Here, we present the {discovery} of a sub-Neptunian mass planet orbiting in the HZ of the nearby mid-M dwarf \hoststar{}. The planet's orbital parameters are derived {from an RV model that includes Keplerian orbits for the planet and another planet candidate, as well as a} red-noise model driven by a Gaussian process (GP) to simultaneously account for the stellar activity.

In \autoref{sec:observations}, we present {the analyzed} ground-based observations. The properties of the host star, \hoststar{}, are introduced in \autoref{sec:HNLib}. The stellar activity analysis and the determination of the rotation period of the parent star are carried out in \autoref{sect:analysis}. Also in this section, with the stellar rotation period constrained, we proceed to analyze the spectroscopic observations with the aim of {identifying} and characterizing planet candidates. The properties of the newly confirmed and discovered planet and the planet candidate orbiting \hoststar{} are given in \autoref{sec:discussion}, together with the discussion on the implications of these findings. Finally, a summary is given in \autoref{sec:summary}.

\section{Observations}
\label{sec:observations}

\begin{table*}
\caption{Available long-term photometric observations of \hoststar{}.}
\label{tab:phot_monitoring}
\centering
\begin{tabular}{lcccSSS}
    \hline \hline
    \noalign{\smallskip}

    Instrument & \multicolumn{2}{c}{Date} & Filter                       & \multicolumn{1}{l}{$\Delta t$\tablefootmark{a} }       & \multicolumn{1}{c}{$N_\textnormal{obs}$\tablefootmark{b}} & \multicolumn{1}{c}{rms\tablefootmark{c}}                                     \\
    
                    & Begin   & End           &                           & \multicolumn{1}{l}{[d]} &  & \multicolumn{1}{c}{[ppt]} \\
    
    \noalign{\smallskip}
    \hline
    \noalign{\smallskip}
    ASAS-3{$^\dagger$}                                      & 24 December 2000         & 13 September 2009            &       $V$                                & 3185                                     & 446                                      & 24.60 \\
    \noalign{\smallskip}
    SuperWASP{$^\dagger$}                                   & 28 February 2010         & 26 July 2010                 &           400--700\,nm                             & 147                                      & 86                                       & 3.23 \\
    \noalign{\smallskip}
    \ldelim\{{4}{*}[\parbox{6em}{ASAS-SN{$^\dagger$}}]              & 27 February 2012         & 10 July 2018                 &          \rdelim\}{4}{17.5mm}[   $V$]                             & 2324                                     & 97                                       & 23.29 \\
                                                & 22 June 2013             & 23 May 2018                  &                                       & 1796                                     & 111                                      & 19.46 \\
                                                & 25 June 2014             & 28 August 2018               &                                       & 1525                                     & 115                                      & 40.24\\
                                                & 10 July 2014             & 28 August 2018               &                                       & 1510                                     & 129                                      & 15.96\\
    \noalign{\smallskip}                                                
    \ldelim\{{4}{*}[\parbox{6em}{MEarth-tel14}]        & 28 May 2017{$^\dagger$}              & 9 June 2017{$^\dagger$}                  &  \rdelim\}{7}{17.5mm}[\   RG715]                                     & 11                                       & 8                                        & 3.72\\
                                                & 1 July 2017{$^\dagger$}              & 8 July 2017{$^\dagger$}                  &                                       & 6                                        & 4                                        & 5.32\\
                                                & 28 August 2017{$^\dagger$}           & 8 September 2017{$^\dagger$}             &                                       & 10                                       & 11                                       & 4.33\\
                                                & 4 June 2019              & 27 February 2022             &                                       & 999                                      & 265                                      & 4.80\\
    \noalign{\smallskip}                                                
    \ldelim\{{3}{*}[\parbox{6em}{MEarth-tel15}]        & 11 June 2017{$^\dagger$}             & 21 June 2017{$^\dagger$}                 &                                       & 9                                        & 7                                        & 2.46\\
                                                & 30 June 2017{$^\dagger$}             & 30 June 2017{$^\dagger$}                 &                                       & 0                                        & 1                                        & 0.98\\
                                                & 13 July 2019           & 16 March 2020                &                                       & 246                                      & 70                                       & 4.10\\
    \noalign{\smallskip}                                                  
    \ldelim\{{2}{*}[\parbox{6em}{OSN-T90}] & 10 May 2018              & 5 June 2019                  & $R$                                   & 391                                      & 49                                       & 8.68\\
                                                & 10 May 2018              & 5 June 2019                  & $V$                                   & 391                                      & 48                                       & 9.00\\
    \noalign{\smallskip}                                                  
    TJO                                         & 9 January 2019           & 9 March 2020                 & $R$                                   & 424                                      & 102                                      & 8.44\\
     \hline
\end{tabular}
\tablefoot{
\tablefoottext{a}{Time-span of the observations.}
\tablefoottext{b}{Binned nightly.}
\tablefoottext{c}{Root mean square in parts-per-thousand.}

Data sets that were not used for the photometric rotational period determination are indicated by a dagger {$^\dagger$}.
}
\end{table*}

\subsection{Spectroscopic data}
\label{subsec:carmenes}

\subsubsection{{CARMENES}}
\hoststar{} was spectroscopically observed with CARMENES between 27 January 2016 and 31 December 2020, resulting in a total of 94 RV measurements. CARMENES is installed at the 3.5\,m telescope of the Calar Alto Observatory in Almer\'ia (Spain). It was specifically designed to deliver high-resolution spectra at optical (resolving power $\mathcal{R} \approx$ 94,600) and near-infrared ($\mathcal{R} \approx$ 80,400) wavelengths covering the range from 520\,nm to 1710\,nm. CARMENES has two different channels, one for the optical (the VIS channel) and one for the near-infrared (the NIR channel) with a dichroic at 960\,nm \citep{Quirrenbach2014}. 
All data were acquired with integration times of 1800\,s, which is the maximum exposure time employed for precise RV measurements {with CARMENES}.

The spectra were processed following the data flow of the CARMENES guaranteed time observations (GTO) program \citep{Ribas2023}. Raw data are reduced with the {\tt caracal} pipeline \citep{Caballero2016b}. Relative RVs are extracted separately for the VIS and NIR channels using the {\tt serval} software \citep{Zechmeister2018}. The final RV per epoch of each channel is computed as the weighted RV mean over all \'echelle orders. The {data from the CARMENES VIS channel are} shown in the bottom panel of \autoref{fig:fit_rv_full} (individual CARMENES relative RVs and their uncertainties are listed in \autoref{tab:hnlib_rv_act_data}). The {root mean square (rms) from the final fit residuals} and the mean errorbars of the CARMENES RV data are {1.75}\,m\,s$^{-1}$ and 1.31\,m\,s$^{-1}${, respectively}.

At high spectral resolution, the profile of the stellar lines may change due to photospheric and chromospheric activity, which has an impact on accurate RV measurements. Therefore, for CARMENES, {\tt serval} provides further measurements for a number of spectral features {in the instruments' wavelength range} that are considered indicators of stellar activity and may have a chromospheric component in active M dwarfs: the differential line width (dLW), the Ca\,{\sc ii} infrared triplet, H$\alpha$, and the chromatic index (CRX). The latter determines the RV--$\log \lambda$ correlation, and it is used as an indicator {for} the presence of stellar active regions. Additionally, as part of the data processing, we also calculate measurements of molecular absorption bands of two species: TiO and VO, together with pseudo-equivalent widths (pEW) for different indices after the subtraction of a reference star spectrum as described by \cite{Schofer2019}. {An overview of the CARMENES activity indicators of \hoststar{} that exhibit significant peaks in the {generalized Lomb-Scargle periodograms \citep[GLS;][]{Zechmeister2009_GLS}} is shown in \autoref{fig:gls_activity}}.

\subsubsection{{Archival data}}

Additional relative {RV} data for \hoststar{} are available in the archives. We employed 14 RVs \citep{Trifonov2020} from the High Accuracy Radial velocity Planet Searcher \citep[HARPS,][]{Mayor2003} at the ESO La Silla 3.6\,m telescope and 34 RV data \citep{Tal-Or2019} from the High Resolution Echelle Spectrometer \citep[HIRES,][]{Vogt1994} on the Keck 10\,m telescope. These velocities span a period of $\sim$9\,yr, almost doubling the total time coverage of our RV analysis. {The mean error bars are 1.47\,m\,s$^{-1}$ for the HARPS data and 2.83\,m\,s$^{-1}$ for HIRES.}

\subsection{{Ground-based} photometry}
There are data from many photometric monitoring campaigns of HN Lib available. An overview of the data used in this work and their properties can be found in \autoref{tab:phot_monitoring}.

\subsubsection{OSN and TJO}
{\hoststar{} was observed by our team using the 90\,cm Ritchey-Chrétien telescope of the Observatorio de Sierra Nevada (OSN) in Almer\'ia, Spain, and the VersArray $\text{2k}\times\text{2k}$ CCD camera \citep{Rodriguez2010}. The total number of frames amounts to 915 in the $V$ filter and 903 in the $R$ filter, where the typical exposure times were 100\,s and 60\,s, respectively. Each filter dataset was corrected for bias and flat fielding before the target flux was determined using differential aperture photometry. Outliers differing by more than $3\sigma$ from the mean were removed from further analysis. The nightly binned light curves are presented in the first two panels of \autoref{fig:fit_photometry}.}

{Further} photometric CCD observations were collected {by us} with the 0.8\,m Telescopi Joan Or\'o \citep[TJO,][]{Colome2010} at the Observatori Astron\`omic del Montsec in Lleida, Spain. A total of 1910 frames with a exposure time of 25\,s were obtained using the Johnson $R$ filter of the LAIA imager, a $\text{4k}\times\text{4k}$ CCD with a field of view of 30\,arcmin, and a scale of 0.4\,arcsec\,pixel$^{-1}$. The images were calibrated with bias, dark, and flat-field frames with the {\tt ICAT} pipeline \citep{Colome2006} of the TJO. The differential photometry was extracted with {\tt AstroImageJ} \citep{Collins2017} using the aperture size that minimized the rms of the resulting relative fluxes, and a selection of the 30 brightest comparison stars in the field that did not show variability. Lastly, we removed outliers and measurements affected by poor observing conditions {with} a low signal-to-noise ratio. The nightly binned TJO $R$-band light curve is presented in the {middle} panel of \autoref{fig:fit_photometry}.

\subsubsection{Archival data}

Photometric public data {of} \hoststar{} were also used for the analysis. {We retrieved archival data from {the MEarth project \citep[][DR 11\footnote{\url{https://lweb.cfa.harvard.edu/MEarth/DR11/README.txt}} -- last two panels of \autoref{fig:fit_photometry}]{Irwin2015}}, the All-Sky Automated Survey \citep[ASAS-3,][]{Pojmanski1997}, the Wide Angle Search for Planets \citep[SuperWASP,][]{Pollacco2006}, and the All-Sky Automated Survey for Supernovae project \citep[ASAS-SN,][]{Shappee2014, Kochanek2017}. \cite{DiezAlonso2019} searched for rotation periods using the ASAS-3 photometric data but did not report a detection for \hoststar{}. To reach a higher precision for our analysis, we created nightly bins of the data from {each instrument} if multiple data points were taken per night.}

{Since the quality of the light curves of ASAS-3 and ASAS-SN is considerably worse than that of the others (cf. \autoref{tab:phot_monitoring}), we did not use them for our final analysis. For the MEarth data, we considered a photometric offset for each version of the respective instruments since they go along with the creation of a new per-star magnitude and meridian offset for the data \citep{Newton2018}. Some resulting sub-data sets have only a few data points and are not very useful due to the ambiguous offset in the fit, which is why we did not consider them in the final analysis (see also \autoref{tab:phot_monitoring}).}


\section{HN Lib}
\label{sec:HNLib}

\hoststar{} is a nearby, single M4.0\,V star located at a distance of only about 6.25\,pc \citep{GaiaCollaboration2022}. Because of its relative brightness, this star has often appeared in the literature and in all-sky surveys \citep[e.g.][]{Schoenfeld1886, Jackson1952, Leggett1992, Mann2015}. 
\autoref{tab:stellar_properties_hnlib} summarises its stellar parameters.

\hoststar{} is cataloged as a variable star that exhibits variations in its luminosity due to rotation of the star coupled with star spots and other chromospheric activity \citep{Kukarkin1982,Weis1994, Hosey2015}. 
Furthermore, \cite{Astudillo-Defru2017} determined a $\log R'_{\rm HK}$ value for \hoststar{} of $-5.58 \pm 0.05$\,dex and estimated from activity-rotation relationships a stellar rotation period of $102 \pm 11$\,d, which classifies our target as a slow rotator. 
This conclusion is in line with the small upper limits of rotational velocity, magnetic field, and X-ray luminosity \citep{Stelzer2013, Reiners2018, Reiners2022}, a new, more robust, mean $\log R'_{\rm HK}$  value of $-5.23 \pm 0.07$\,dex homogeneously computed by averaging individual values collected with HARPS, FEROS, HIRES, and ESPaDONS \citep{Perdelwitz2021}, the measurement of chromospheric lines in absorption or moderate emission \citep{Fuhrmeister2020}, and our own rotation period determination (see below).

The fundamental atmospheric parameters ({i.e.,} $T_{\rm eff}$, $\log{g}$, and [Fe/H]) of \hoststar{} using the CARMENES VIS and NIR template spectra were recently determined by \cite{Marfil2021} by means of the {\sc SteParSyn} code\footnote{\url{https://github.com/hmtabernero/SteParSyn/}} \citep{Tabernero2022}. The resulting values, listed in \autoref{tab:stellar_properties_hnlib}, agree at the $2\sigma$ level with those derived by other authors via spectral synthesis \citep[e.g.][]{Rojas-Ayala2012, Gaidos2015, Passegger2018, Rajpurohit2018, Maldonado2020} and from fits to the photometric spectral energy distribution \citep{Cifuentes2020}.

We followed the recipes of the latter authors to determine a new bolometric luminosity, $L_\star$, with updated {\em Gaia} DR3 data.
We used $L_\star$ and $T_{\rm eff}$ to determine the mass and radius of \hoststar{} {to be $M_\star = 0.291 \pm 0.014$\,M$_\odot$ and $R_\star = 0.299 \pm 0.009$\,R$_\odot$} with Stefan–Boltzmann law and the mass-radius relation of \cite{Schweitzer2019}, which was based on detached, double-lined, double-eclipsing, main-sequence M-dwarf binaries from the literature. 

Besides, we employed the spectral synthesis method, together with the PHOENIX BT-Settl atmospheric models \citep{Allard2012} and the radiative transfer code {\tt Turbospectrum} \citep{Plez2012} for determining Mg and Si abundances of \hoststar{}. We measured [Mg/H] = $-0.16 \pm 0.14$\,dex and [Si/H] = $0.06 \pm 0.20$\,dex. Further details on the followed procedure will be provided by Tabernero et~al. (in~prep.).

\begin{table}
\centering
\small
\caption{Stellar parameters of HN~Lib.} 
\label{tab:stellar_properties_hnlib}
\begin{tabular}{lcr}
\hline
\hline
\noalign{\smallskip}
Parameter & Value & Reference \\ 
\noalign{\smallskip}
\hline
\noalign{\smallskip}
\multicolumn{3}{c}{\em Basic identifiers and data}\\
\noalign{\smallskip}
BD                                  & --11 3759             & Scho86 \\ 
Wolf                                & 1481                  & Wol25 \\
Gl                                  & 555                   & Gli69 \\
Karmn                               & J14342--125           & AF15, Cab16b \\
Sp. type                            & M4.0\,V      & Haw96 \\
$G$ [mag]                           & $9.8950 \pm 0.0030$ & {\it Gaia} DR3$^a$ \\
$J$ [mag]                           & $6.838 \pm 0.019$ & 2MASS$^a$ \\
\noalign{\smallskip}
\multicolumn{3}{c}{\em Astrometry and kinematics}\\
\noalign{\smallskip}
$\alpha$ (ICRS, epoch 2016.0)                    &  14:34:16.424  & {\it Gaia} DR3 \\
$\delta$ (ICRS, epoch 2016.0)                    & --12:31:00.92  & {\it Gaia} DR3 \\
$\mu_{\alpha}\cos\delta$ [$\mathrm{mas\,yr^{-1}}$]  & $-355.138 \pm 0.053$ & {\it Gaia} DR3 \\
$\mu_{\delta}$ [$\mathrm{mas\,yr^{-1}}$] & $+593.040 \pm 0.044$ & {\it Gaia} DR3 \\
$\varpi$ [mas]                      & $159.923 \pm 0.055$ & {\it Gaia} DR3 \\ 
$d$ [pc]                            & $6.2530 \pm 0.0022$ & {\it Gaia} DR3 \\ 
$\gamma$ [$\mathrm{km\,s^{-1}}$]    & $-1.426 \pm 0.006$ & Sou18 \\ 
$\dot{\gamma}$ [$\mathrm{m\,s^{-1}\,yr^{-1}}$] & $+0.06901 \pm 0.00041$ & This work \\
$U$ [$\mathrm{km\,s^{-1}}$]         & $-13.8606 \pm 0.0051$ & This work \\
$V$ [$\mathrm{km\,s^{-1}}$]         & $+5.8821 \pm 0.0053$ & This work \\
$W$ [$\mathrm{km\,s^{-1}}$]         & $+13.9700 \pm 0.057$ & This work \\
Galactic population                 & Thin disk & This work \\
\noalign{\smallskip}
\multicolumn{3}{c}{\em Fundamental parameters and abundances}\\
\noalign{\smallskip}
$T_{\mathrm{eff}}$ [K]              & $3347 \pm 50$     & Mar21 \\ 
$\log{g}$                           & $4.76 \pm 0.13$   & Mar21 \\
{[Fe/H]}                            & $-0.18 \pm 0.15$  & Mar21 \\
{[Mg/H]}                            & $-0.16 \pm 0.14$  & Tab \\
{[Si/H]}                            & $+0.06 \pm 0.20$  & Tab \\
$L_\star$ [$\rm 10^{-6}\,L_\odot$]      & $10106 \pm 69$    & This work \\
$R_\star$ [R$_{\odot}$]             & $0.299 \pm 0.009$ & This work \\
$M_\star$ [M$_{\odot}$]             & $0.291 \pm 0.013$ & This work \\
\noalign{\smallskip}
\multicolumn{3}{c}{\em Activity and age}\\
\noalign{\smallskip}
$v \sin i_\star$ [$\mathrm{km\,s^{-1}}$] & $<2.0$       & Rei18 \\
$P_{\rm rot}$ [d]                   & $96 \pm 2$            & This work \\ 
pEW(He~{\sc i} D$_3$) [\AA]         & $+0.070 \pm 0.010$ & Fuh20 \\
pEW(H$\alpha$) [\AA]                & $+0.088 \pm 0.037$ & Fuh20 \\ 
pEW(Ca~{\sc ii} IRT$_1$) [\AA]      & $-0.006 \pm 0.015$ & Fuh20 \\
pEW(He~{\sc i} IR) [\AA]            & $+0.498 \pm 0.004$ & Fuh20 \\
$\log R'_{\rm HK}$                  & $-5.226^{+0.063}_{-0.074}$ & This work$^{b}$ \\ 
$\langle B \rangle$ [G]             & $<$ 220 & Rei22 \\ 
$\log{L_{\rm X}}$ [erg\,s$^{-1}$]   & $<$ 26.80 & This work$^c$ \\ 
Age [Gyr]                           & 0.8--8.0 & This work \\
\noalign{\smallskip}
\hline
\end{tabular}
\tablebib{
2MASS: \citet{Skrutskie2006};
AF15: \citet{AlonsoFloriano2015};
Cab16b: \citet{Caballero2016a};
Fuh20: \citet{Fuhrmeister2020};
{\em Gaia} DR3: \citet{GaiaCollaboration2021};
Gli69: \citet{Gliese1969};
Mar21: \citet{Marfil2021};
Haw96: \citet{Hawley1996};
Rei18: \citet{Reiners2018};
Rei22: \citet{Reiners2022};
Scho86: \citet{Schoenfeld1886};
Ste13: \citet{Stelzer2013};
Sou18: \citet{Soubiran2018};
Tab: Tabernero et~al. (in prep.);
Wol25: \citet{Wolf1925}.
}
\tablefoot{
\tablefoottext{a}{See \citet{Cifuentes2020} for multiband photometry different from {\em Gaia} $G$ and 2MASS $J$.}
\tablefoottext{b}{From data compiled by \citet{Perdelwitz2021}.}
\tablefoottext{c}{Computed by us from a $\log{f_{\rm X}}$ upper limit by \citet{Stelzer2013}.}
}
\end{table}

The galactic space velocities $UVW$ of \hoststar{} were derived using the $Gaia$ DR3 coordinates, proper motions, and systemic radial velocity with the formulation developed by \cite{Johnson1987}.

\hoststar{} does not appear to belong to any known young stellar moving group and has kinematics typical of a stars in the galactic thin disk, indicating a likely age of 0.8--8.0\,Gyr, which is in agreement with the very weak stellar activity.
The proximity of \hoststar{} has made our target to be the subject of numerous searches for close companions \citep[e.g.,][]{Jameson1983, Skrutskie1989, Nakajima1994, Simons1996, Oppenheimer2001, Hinz2002, Tanner2010, Jodar2013, Ward-Duong2015, Davison2015, CortesContreras2017}. The high-resolution images of \cite{Dieterich2012} were the most sensitive to very-low-mass
stars, brown dwarfs, or planets using NICMOS on the \textit{Hubble} with the F180M near-infrared filter. \cite{Dieterich2012} established companion magnitude and angular separation limits ranging from 11.5\,mag at 0.4\,arsec (2.5\,au) to 18\,mag at 3.0-4.0\,arcsec (19--25\,au) for companions that could be ruled out for \hoststar{}. These limits, together with the age of the system and the evolutionary models for very-low-mass stars and brown dwarfs with dusty atmospheres from \cite{Baraffe2015}, discard any hypothetical companion with mass from 0.07 to 0.08\,M$_{\odot}$ at projected orbital separations larger than 0.4\,arsec, and from 0.025 to 0.06\,M$_{\odot}$ at separations larger than 3.0-4.0\,arcsec.
Finally, as in \citep{Caballero2022}, we searched for objects with common {\em Gaia} DR3 parallax and proper motions up to a projected physical separation of \num{100000}\,au.
We found no hints of any wide potential companion.

\section{Analysis and results}
\label{sect:analysis}

\subsection{Modeling techniques}

To get a quick overview of all the different data available to us, we used the graphical user interface capabilities of \texttt{Exo-Striker}\footnote{\url{https://github.com/3fon3fonov/exostriker}} \citep{Trifonov2019} to generate GLS periodograms of RVs, photometry, and spectral activity indicators. 
For the retrieval of the parameters presented in the final analysis of this work, we used however the \texttt{juliet}\footnote{\url{https://juliet.readthedocs.io/en/latest/index.html}} package \citep{Espinoza2019} because of the variety of GP regression kernels that it provides. It brings together frequently used tools {such as} \texttt{radvel} \citep{Fulton2018} for RV modeling and \texttt{celerite} \citep{Foreman-Mackey2017} and \texttt{george} \citep{Ambikasaran2015} for the implementation of GPs. For the fitting process, \texttt{juliet} offers different options. We chose to use \texttt{dynesty} \citep{Speagle2020}, which implements a nested sampling algorithm and {supports} computing Bayesian log evidence, $\ln\mathcal{Z}$, for the models.

\subsection{Stellar activity and rotation period}
\label{subsec:rotation}

\subsubsection{Activity indicators}
\begin{figure*}[ht]
\center
\includegraphics[width=\columnwidth]{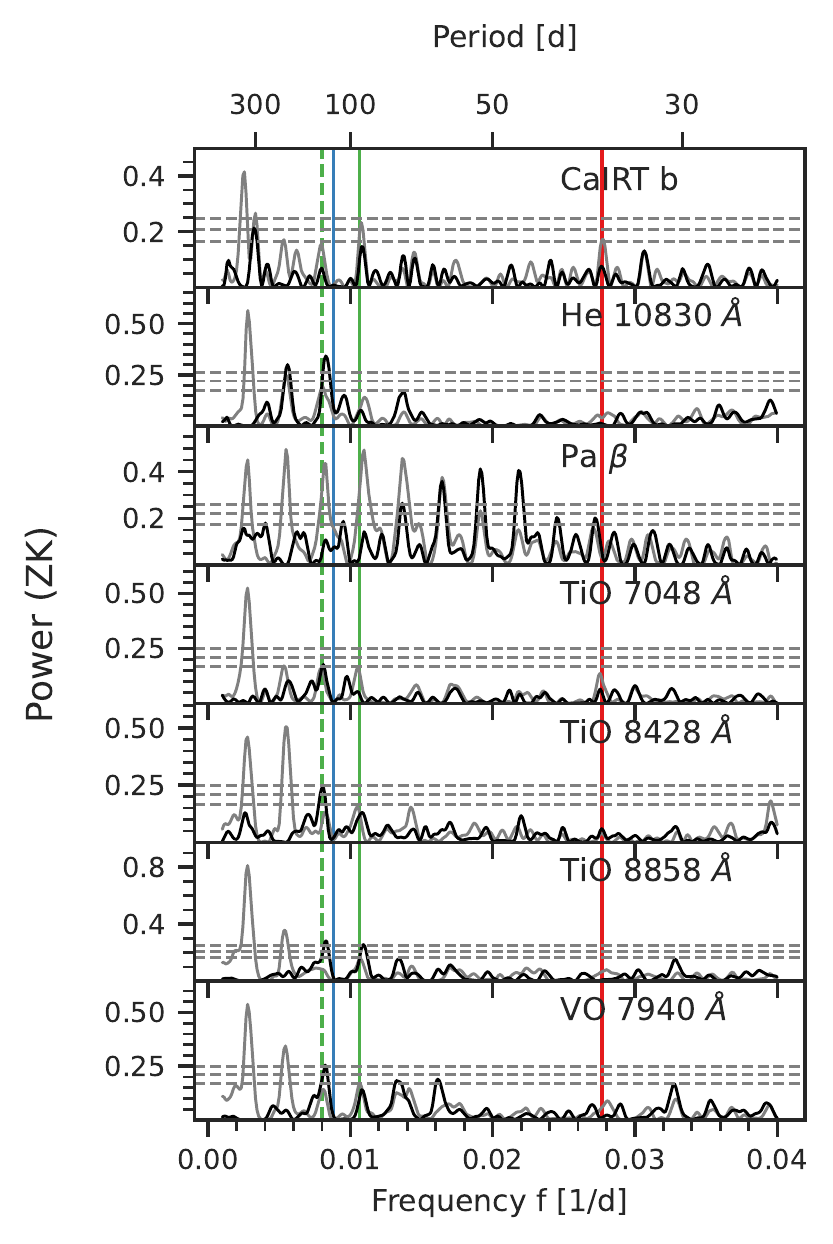}
\includegraphics[width=\columnwidth]{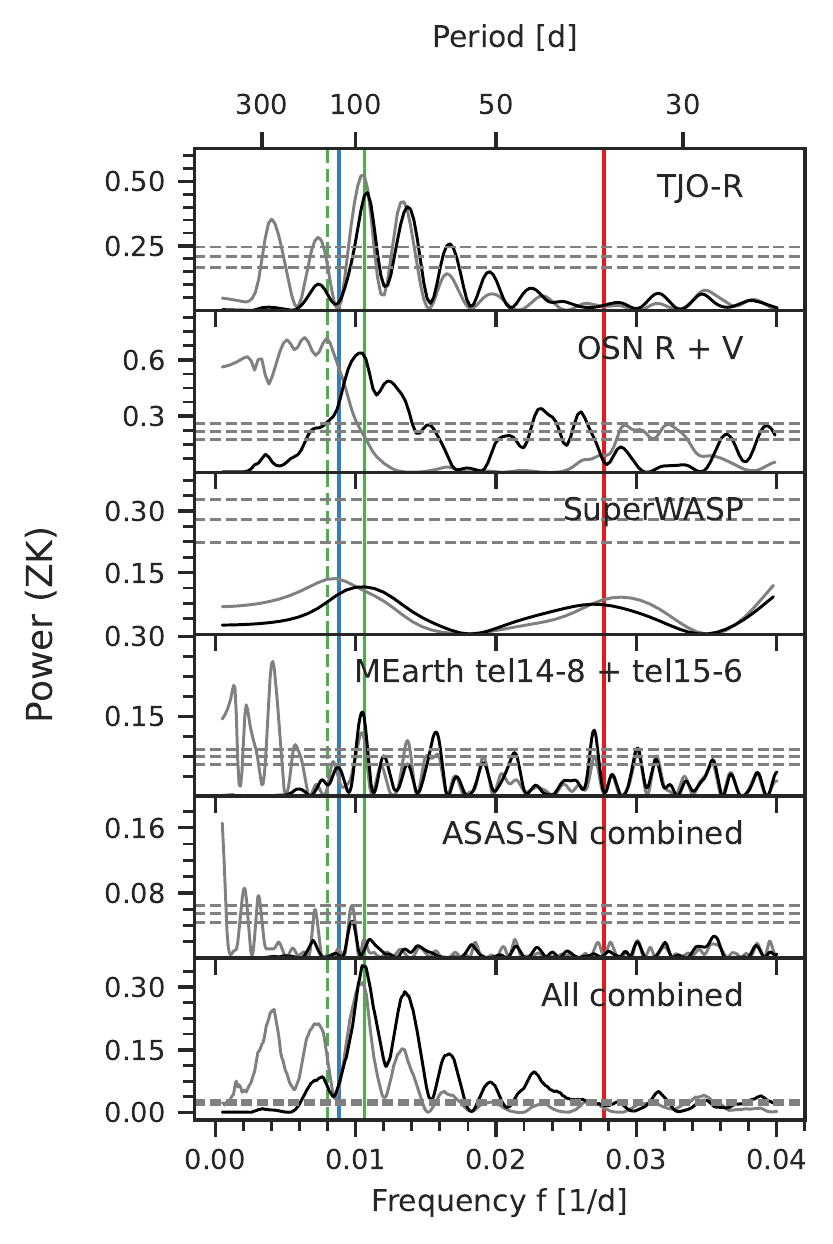}
\caption{Signal search with GLS periodograms in the spectral activity indicators ({\em left}) {and photometric} data ({\em right}) of \hoststar{}. The data of the activity indicators were pre-whitened by subtracting the long-period $\sim$360\,d or $\sim$180\,d signals accordingly, whereas the photometric data were detrended using {\tt CAFM} to remove the appearing long-term trends. Original periodograms are shown in grey and the corrected ones in black. We mark the \Pbday{} signal {from the RVs} with a red solid line and the {\Pcday{} period} with a blue solid line. The common alias pair of the activity indicators with periods of \Protaliasone{} and \Protaliastwo{} are depicted by green dashed and solid lines, respectively. FAPs of 10, 1, and \SI{0.1}{\percent} were calculated with the analytical expression of \cite{Zechmeister2009_GLS} and are shown with the horizontal grey dashed lines {from bottom to top}.}\label{fig:gls_activity}
\end{figure*}

\begin{table}[]
\centering
\caption{Derived stellar rotation period values using various methods.}
\label{tab:summary_Prot}
\begin{tabular}[t]{lc}
\hline
\hline
\noalign{\smallskip}
Method & Value \\
\hline
\noalign{\smallskip}
$\log {R'}_{\rm HK}^{(a)}$      & $102 \pm 11$\,d \\
Spectral act. indicators$^{(b)}$ & 94 and 125 \,d \\
Phot. QP-GP kernel               & $94 \pm 4$\,d \\
Phot. dSHO-GP kernel             & $96 \pm 2$\,d \\
\noalign{\smallskip}
\hline
\end{tabular}  
\tablefoot{
\tablefoottext{a}{\cite{Astudillo-Defru2017}.}
\tablefoottext{b}{These values are aliases one of each other.} 
}
\end{table}

To identify signals that originate from stellar activity, we investigated the GLS periodograms for the range of activity indicators determined from the CARMENES VIS and NIR spectra as described in \autoref{subsec:carmenes}. Since our observations {span} several years, many of the periodograms suffer from strong signals with periods of 360\,d to 380\,d, as well as approximately half a year, caused by {uncorrected contamination by telluric lines}. {For those spectroscopic activity indicators that show peaks at these periodicities,} we proceeded with periodograms that were pre-whitened by subtracting the best matching solution to these signals from the data. For clarity, we show only the periodograms for the activity indicators with peaks of less than \SI{10}{\percent} false alarm probability (FAP) after the pre-whitening in \autoref{fig:gls_activity}. All shown indicators share a pair of signals with periods of \Protaliasone{} and \Protaliastwo{} that are caused by aliasing due to the seasonal observability of \hoststar{}\footnote{The seasonal observability imprints a signal of about ~1/365\,d$^{-1}$ to the window function of the data. Aliasing occurs at $f_a=f_t\pm m \times f_s$, where $f_a$ are the alias frequencies, $f_t$ is the true underlying signal, $m$ is an integer value and $f_s$ is the sampling frequency. Given a signal of \Protaliasone{}, we therefore expect first-order aliasing at $f_a=1/125\text{\,d}+1/365\text{\,d} \approx 1/93\text{\,d}$.}. Both periods are close to the {estimate} of $P_\text{rot,est}=102\pm11$\,d based on the $\log R'_{\rm HK}$ measurement by \cite{Astudillo-Defru2017}, which suggests that the underlying signal is indeed related to the stellar rotation. However, since both {signals} are within the tolerance of the predicted rotation period, we could not resolve the aliasing {with the spectroscopic data}.

\subsubsection{Photometry}

\begin{figure}
\centering
\includegraphics[width=\columnwidth]{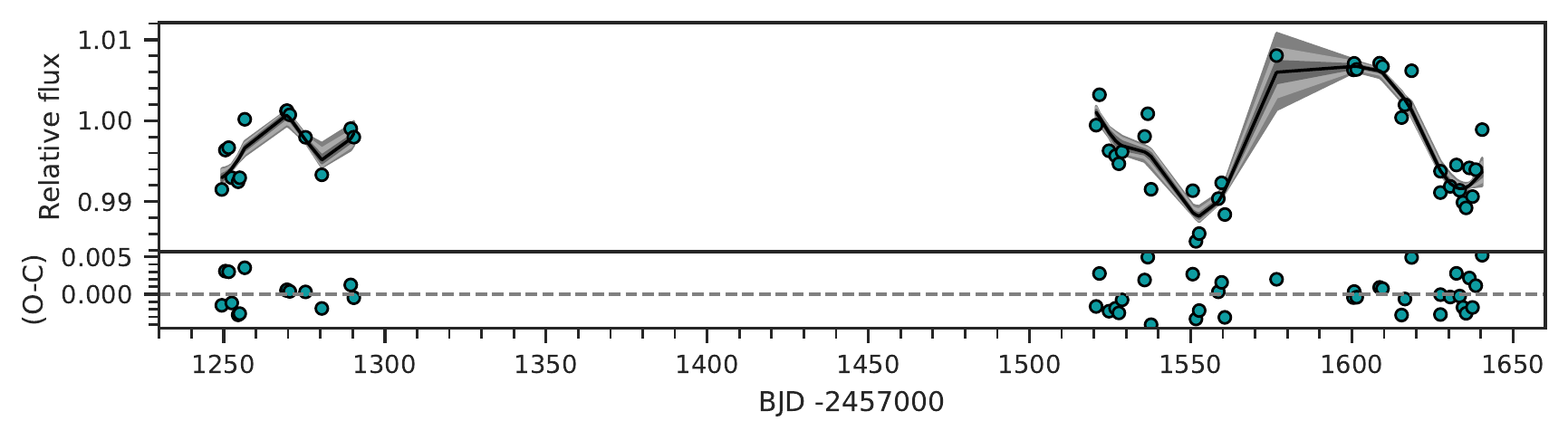}\\
\includegraphics[width=\columnwidth]{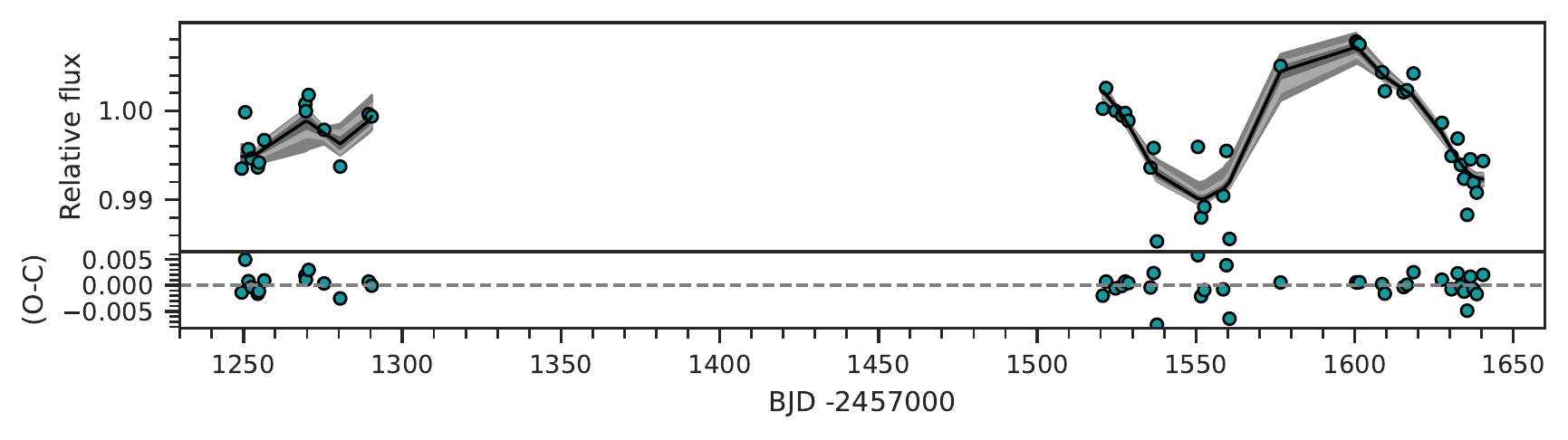}\\
\includegraphics[width=\columnwidth]{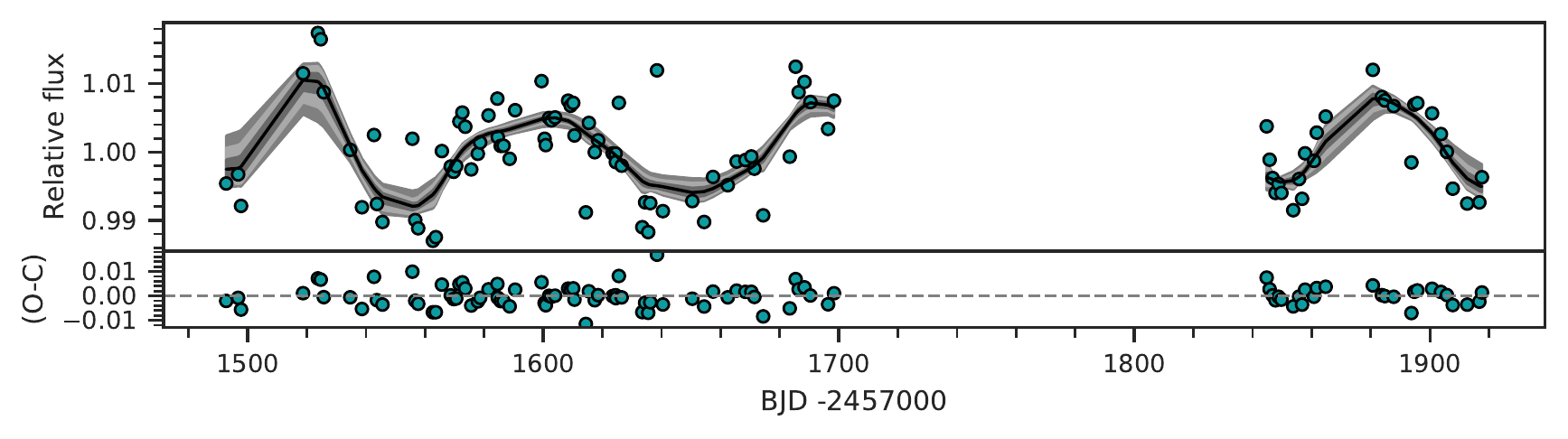}\\
\includegraphics[width=\columnwidth]{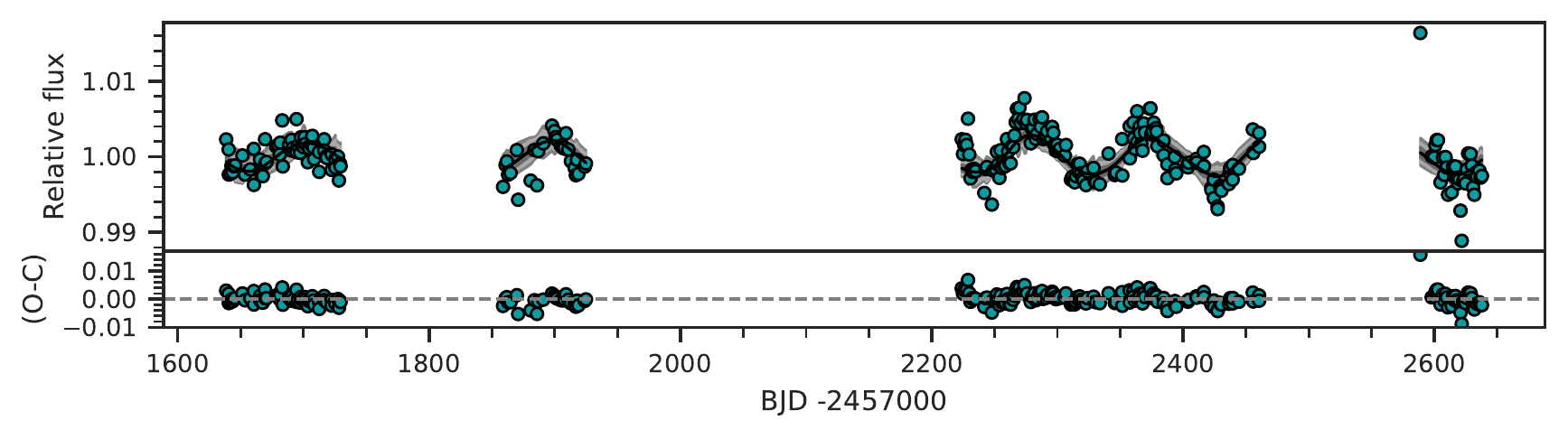}\\
\includegraphics[width=\columnwidth]{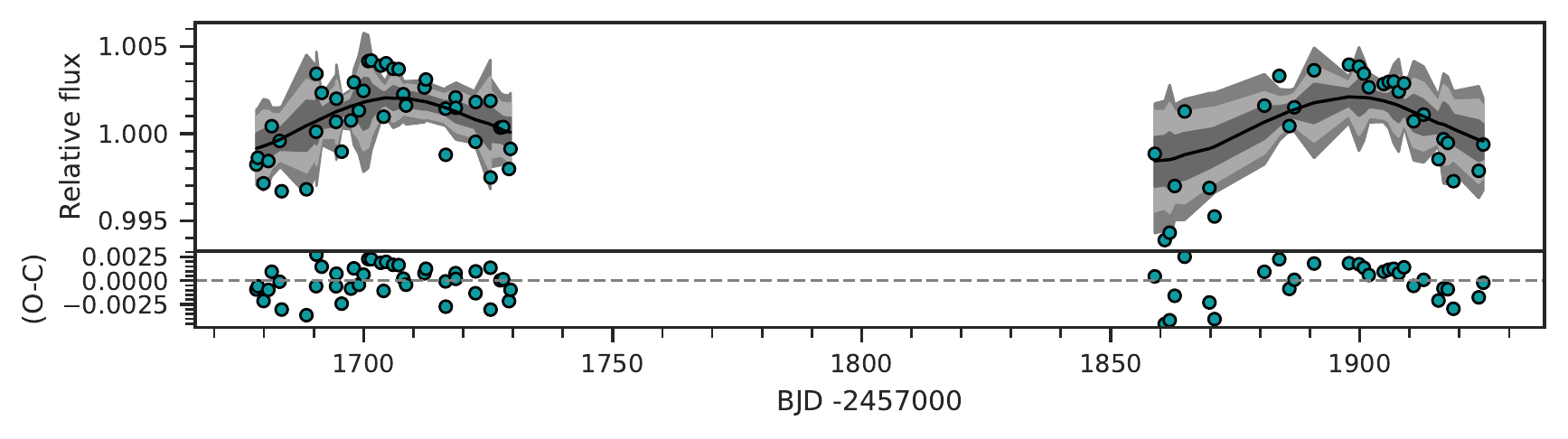}
\caption{Photometric time series of \hoststar. The median models of the GP fit from \num{10000} samples from the posterior are shown by the black lines and the {grey} shaded areas denote the \SI{68}{\percent}, \SI{95}{\percent,} and \SI{99}{\percent} {confidence} intervals, respectively. \textit{From top to bottom}: {OSN-R, OSN-V, TJO, MEarth-tel14, {and MEarth-tel15}.}} 
\label{fig:fit_photometry}
\end{figure}

The GLS periodogram of the {TJO photometric time series} in \autoref{fig:gls_activity} shows a peak at 95\,d, with aliases at 74\,d and 135\,d due to the seasonal observations. Furthermore, a long periodic signal with a period of 263\,d is visible, that could be an alias of a longer activity cycle of the star. 
Because of the data gap of about $\sim$200\,d (and the consequently uncertain offset) between the two OSN observing runs and the short baseline of each data chunk, the GLS periodogram of the OSN data shows a large plateau for long periods. However, when inspected visually, the OSN and TJO photometric times series pretty much overlap (see \autoref{fig:comparison_TJO-OSN} for comparison). A periodicity of $\gtrsim100$\,d is  also apparent in the periodogram of the daily binned data from the SuperWASP survey; however, despite being the most prominent peak, it is not significant in the GLS. A signal of an unknown origin at a period of 247\,d is dominant in the MEarth data. {In the regime of around $100$\,d, the strongest peak appears at 96\,d, which is consistent with the TJO data.

Based on the $\log R'_{\rm HK}$ measurements and the GLS analysis of the activity indicators, we expected the rotation period to be in the range between 90 and 130 days. Consequently, we detrended the data before further analysis due to the appearing long-term trends. For this, we used \texttt{wotan} \citep{Hippke2019} and Cosine Filtering with Autocorrelation Minimization \citep[{\tt CFAM} --][]{Kipping2013, Rodenbeck2018}, where we set the window length to 200\,d. The CFAM had a particularly strong effect on the OSN data, as it removed the offset between the two observing seasons. Overall, the detrended photometric data share a common periodicity, which appears as a very significant signal in the combined GLS periodogram at a period of 97\,d, with yearly aliases of first and second order {on both sides.}}

To measure the photometric rotation period of \hoststar{} considering all photometric data combined, we used GP modeling. In contrast to the GLS periodograms, which {are based} on static sinusoidal models, GPs are able to represent also the quasi-periodic nature of stellar activity. For our purposes, we used two different kernels. The first one was a kernel that is the sum of two simple harmonic oscillators, in the following abbreviated as ``dSHO'' kernel \citep[e.g.][]{David2019, Gillen2020}. A description of the implementation and parametrization of this kernel in \texttt{juliet} can be found in \cite{Kossakowski2021}. The second kernel was the multiplication of an exponential-sine-squared kernel with a squared-exponential kernel, which is usually called the ``quasi-periodic'' (QP) kernel \citep{Haywood2014, Rajpaul2015}. The parametrization in \texttt{juliet} for this kernel was described by \cite{Espinoza2019}.

{Because they are far away in time from the MEarth, {OSN}, and TJO observations and, thus, also from our RVs, we did not include the {SuperWASP} data in the fits, considering that the star could have been in a different stage of the activity cycle \citep{Baliunas1995, DiezAlonso2019, Fuhrmeister2023}.} 
We used shared hyperparameters between the instruments for the rotation period and the parameters that describe the coherence and shape of the periodic components of the kernels (i.e., the quality factor and the difference in the quality factor of the dSHO kernel and the $\alpha$ and $\Gamma$ parameters for the quasi-periodic kernel). The amplitude of the GP was considered individually for each instrument to account for potential wavelength {dependencies} and different states of activity. A summary of the GP hyperparameters and the used priors can be found in \autoref{tab:priors_phot}. The
MEarth data used comparison stars that are not M dwarfs \citep{Irwin2011, Newton2016}, introducing systematics. Their light curves were corrected by applying a simultaneous linear detrending of these data using the common mode parameter, which is based on the observations of all M-dwarf targets in the MEarth data combined to make a lower cadence (binned) comparison light curve \citep{Newton2016, Newton2018}.

Our fit with uninformative priors on the period between 10\,d and 200\,d resulted in a period of {$94 \pm 4$\,d using the QP-GP and $96 \pm 2$\,d using the dSHO-GP.} The photometry and the model from the dSHO-GP fit are shown in \autoref{fig:fit_photometry}.
{The period from the photometric fit hence matches the \Protaliastwo{} period of the spectroscopic activity indicators (see the left panels of Fig.~\ref{fig:gls_activity}), although it is the least significant of the two occurring aliases in most of the non-pre-whitened periodograms. One explanation for it could be that both \Protaliasone{} and \Protaliastwo{} are close to harmonics of the long-term signals with periods of approximately one year, which we removed by pre-whitening the data. 
Since \Protaliasone{} matches a lower order harmonic, it is not unexpected that it is also more affected by residual signal in the data and thus increased in power in the GLS. In agreement with the $\log R'_{\rm HK}$ measurements and the spectral activity indicators, we therefore established a rotation period of $96 \pm 2$\,d for \hoststar{}, the more precise value obtained from the dSHO-GP fit. In Table~\ref{tab:summary_Prot} we summarize the different values derived for the rotation period of the star using various methods (i.e., activity-rotation relationships, CARMENES spectroscopic activity indices, and photometric light curves) to ensure that all of them are consistent within their error bars. The error bars for the photometric cases come from our GP fit.}

\subsection{RV signal detection}

\begin{figure}
\center
\includegraphics[width=\columnwidth]{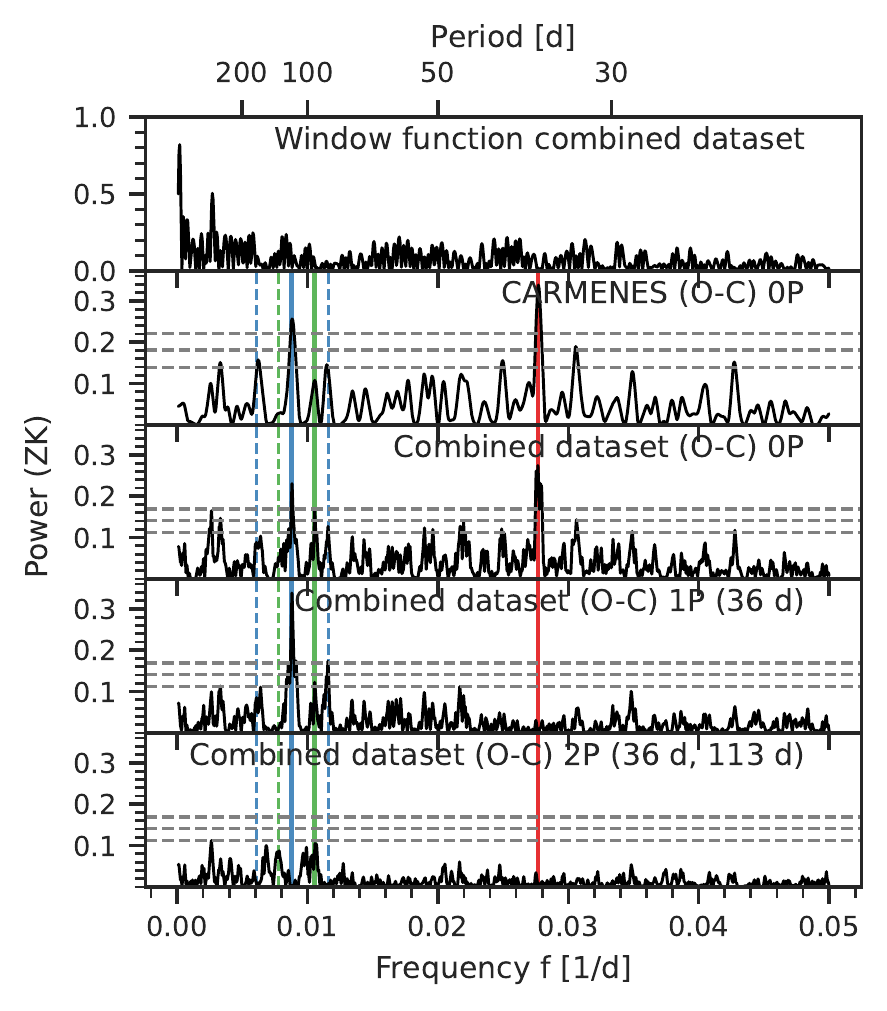}
\caption{Signal search with GLS periodograms in the RV data of \hoststar{}. The first panel shows the window function of the combined CARMENES, HIRES, and HARPS dataset. In the second panel, we show the GLS of only the CARMENES data. The subsequent panels present the GLS periodograms of the combined data. Each panel shows the residuals after subtracting models of increasing complexity. Inset texts {describe} the type of model that was applied to the data before the GLS was generated. A red solid line marks the period of the \Pbday{} signal {and the blue line shows the \Pcday{} signal, while the green line denotes the rotation period of \hoststar{}}. {The 365\,d alias frequencies of the \Pcday{} signal and the stellar rotation period are highlighted by dashed lines.} FAPs of 10, 1, and \SI{0.1}{\percent} were calculated, using the analytical expression of \cite{Zechmeister2009_GLS} and are shown by the horizontal gray dashed lines.}
\label{fig:gls_rv}
\end{figure}

\begin{figure*}[]
        \centering
        \includegraphics{legend_rv_plots.pdf}\vspace{-1.0em}
        \includegraphics[width=0.48\textwidth]{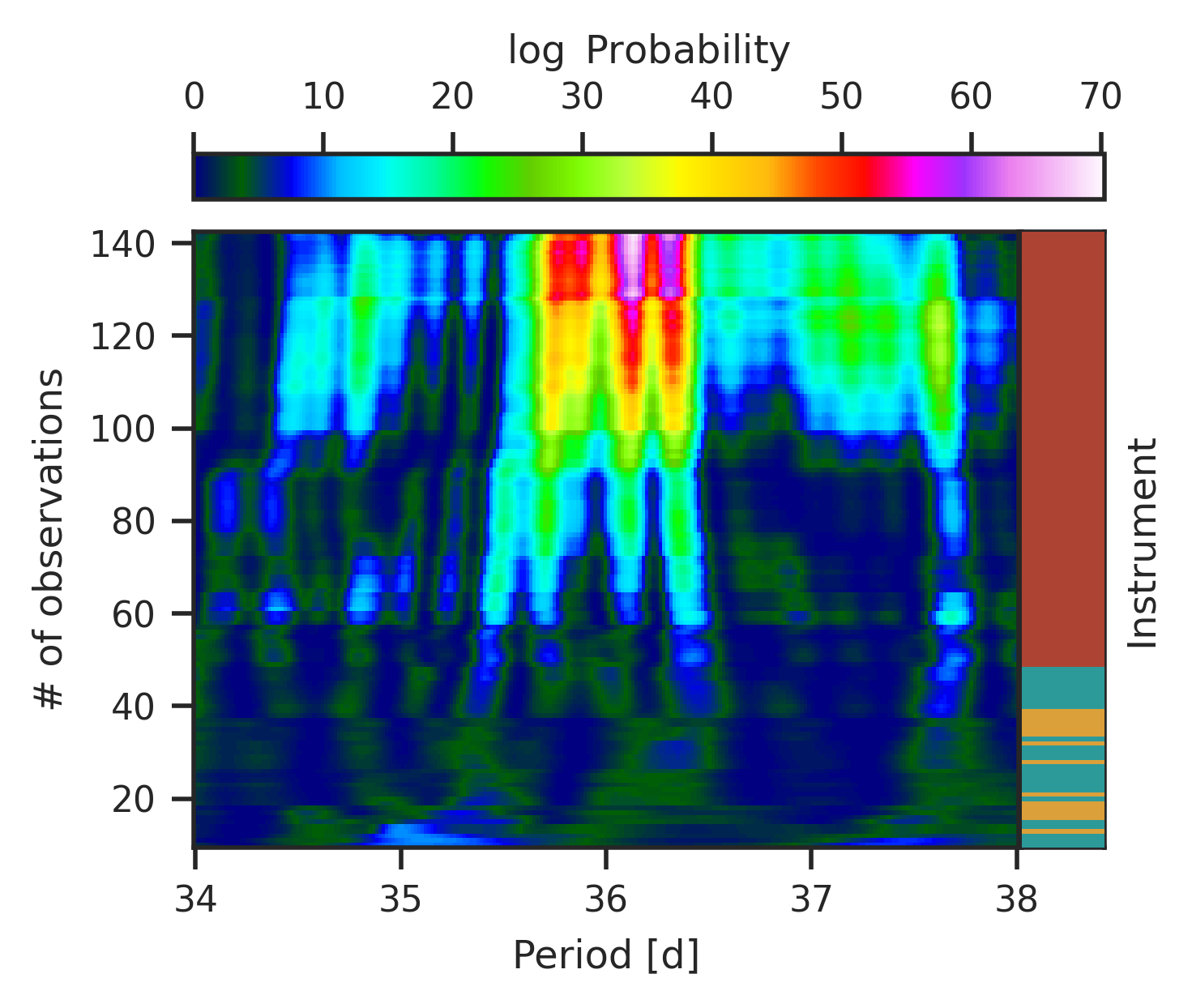}
        \includegraphics[width=0.48\textwidth]{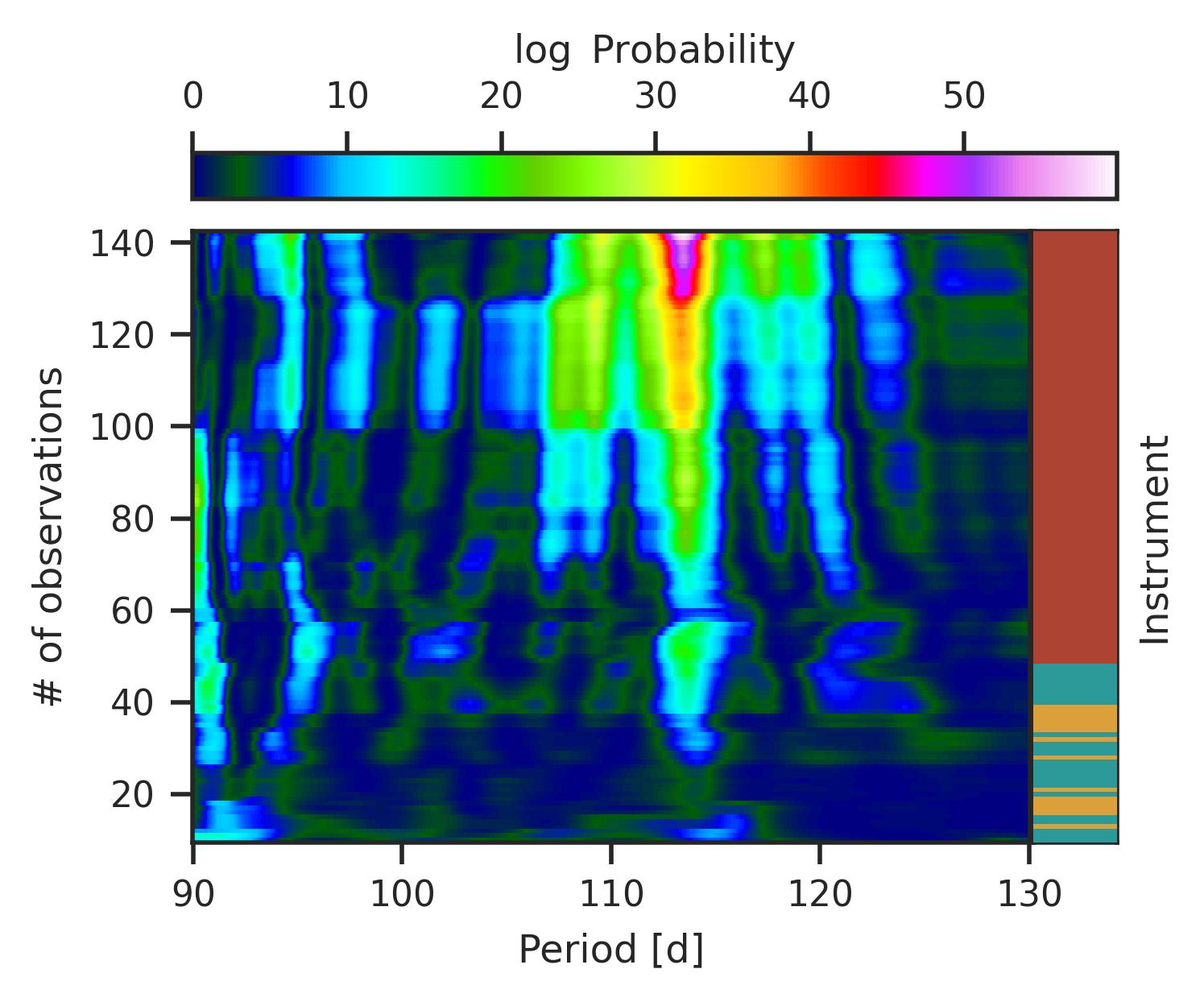}
        \caption{Evolution of the s-BGLS periodogram of the CARMENES, HIRES, and HARPS RV data around 36\,d ({\em left}) and in the region between 90\,d and 130\,d ({\em right}). The periodograms were generated using the residuals after subtracting the respective other periodicity, namely, the 113-d period in the left panel and the 36-d signal in the right panel. Both s-BGLS periodograms include the stellar rotation period. The colorbar on the top shows the Bayesian probability, where higher values are more likely. The right side of each plot indicates the instrument from which the individual added data points originate.}
        \label{fig:hnlib_color_s-BGLS}
\end{figure*}

We began the analysis of the RV data with a signal search using the GLS periodograms, as shown in \autoref{fig:gls_rv}. {We considered a signal to be significant in the GLS periodogram if it has an FAP $<$ \SI{1}{\percent}}. The CARMENES data (second row {of \autoref{fig:gls_rv}}) show two significant peaks at periods of $\sim$36.1\,d and $\sim$113.1\,d. When combining the CARMENES data with the HIRES and HARPS measurements (third row and following), both signals remain significant. However, {multiple alias peaks appear on both sides of the \Pbday{} signal} due to the long {time} baseline and seasonal sampling of the combined data. {Considering that this aliasing does not occur in the CARMENES data alone}, we were {consequently} confident in the $\sim$36.1\,d period of the \Pbday{} signal. {Further, a signal with a FAP of almost $0.1\%$ is visible near the period of the stellar rotation period (third row of \autoref{fig:gls_rv}), as determined in the previous section. The peak is distinct from the \Pcday{} signal and not related by aliasing, as shown by the dashed lines in \autoref{fig:gls_rv}.} Subtracting the {\Pbday{} signal} with a sinusoidal model considerably increases the power of the \Pcday{} signal in the residuals (fourth row of \autoref{fig:gls_rv}){, but decreases the power of the signal at the rotation period.} The residuals of a simultaneous sinusoidal fit of the \Pbday{} signal and the \Pcday{} period do not show any further significant signals (fifth row {of \autoref{fig:gls_rv}}). Since the period of $\sim$113.1\,d {is close to} the rotation period determined in \autoref{subsec:rotation}, we assumed {that it could be} an imprint of the stellar activity onto the RVs {and, consequently,  performed a more in-depth analysis of this signal, which is presented in \autoref{subsec:113d}}.

We verified that {both significant} RV signals {are} stable and coherent over the entire observational time baseline by producing the stacked Bayesian generalized Lomb-Scargle periodograms \citep[s-BGLS,][]{Mortier2015} shown in \autoref{fig:hnlib_color_s-BGLS}. The probability of the {\Pbday{}} signal increases with time until {a constant high degree} is reached above one hundred observations. The aliases of 1\,yr are also seen on both sides of the 36.1\,d {period, but with varying probability. With the increasing number of observations, the uncertainty of the period of the signal also} becomes narrower, which is a behavior expected for signals with a Keplerian origin \citep{Stock2020, Bluhm2021, Chaturvedi2022}. Remarkably, the \Pcday{} signal, {for which we} expected a more erratic behavior in case of a stellar activity origin, seems to be also a stable signal after reaching $\sim$ 80 observations. The region between 94\,d and 96\,d has an unstable behaviour revealed by changes in its width and probability when increasing the number of observations. As a result, it is likely to the site of stellar activity.

\subsection{RV modeling and planetary parameters}
\label{subsec:rvmodelling}

{\setlength{\extrarowheight}{4.5pt}
\begin{table*}[!ht]
\caption{Bayesian model comparison for the RV data.
}
\label{tab:modelcomp}
\begin{tabular}{l p{3cm} p{2cm} S S S}
\hline
\hline
                            Model &                                                                     Priors [d] &                             Posteriors [d] &  $\ln\mathcal{Z}$ &  $\Delta\ln\mathcal{Z}$ &  {max. log likelihood} \\
\noalign{\smallskip}
\hline
                     0P + dSHO$_\text{(96 d)}$ &                                                                                 $\mathcal{N}_\mathrm{Prot, GP}(96,6)$ &                                94.9273 &            -397.7 &                   -15.0 &        -378.4 \\
             1P$_\text{(36 d-ecc)}$ + dSHO$_\text{(96 d)}$ &                                          $\mathcal{U}_\mathrm{b}(30,40)$\newline$\mathcal{N}_\mathrm{Prot, GP}(96,6)$ &                36.1234\newline103.1397 &            -388.7 &                    -6.0 &        -351.6 \\
            1P$_\text{(36 d-circ)}$ + dSHO$_\text{(96 d)}$ &                                          $\mathcal{U}_\mathrm{b}(30,40)$\newline$\mathcal{N}_\mathrm{Prot, GP}(96,6)$ &                36.1185\newline101.1691 &            -388.6 &                    -5.9 &        -355.4 \\
  2P$_\text{(36 d-ecc, 113 d-ecc)}$ + dSHO$_\text{(96 d)}$ & $\mathcal{U}_\mathrm{b}(30,40)$\newline$\mathcal{U}_\mathrm{c}(100,120)$\newline$\mathcal{N}_\mathrm{Prot, GP}(96,6)$ & 36.1154\newline113.4702\newline98.3681 &            -387.1 &                    -4.4 &        -344.0 \\
 2P$_\text{(36 d-circ, 113 d-ecc)}$ + dSHO$_\text{(96 d)}$ & $\mathcal{U}_\mathrm{b}(30,40)$\newline$\mathcal{U}_\mathrm{c}(100,120)$\newline$\mathcal{N}_\mathrm{Prot, GP}(96,6)$ & 36.1234\newline113.4132\newline98.3081 &            -384.5 &                    -1.8 &        -341.6 \\
2P$_\text{(36 d-circ, 113 d-circ)}$ + dSHO$_\text{(96 d)}$ & $\mathcal{U}_\mathrm{b}(30,40)$\newline$\mathcal{U}_\mathrm{c}(100,120)$\newline$\mathcal{N}_\mathrm{Prot, GP}(96,6)$ & 36.1173\newline113.4677\newline99.2176 &            -382.9 &                    -0.2 &        -344.1 \\
 {\bf 2P$_\text{(36 d-ecc, 113 d-circ)}$ + dSHO$_\text{(96 d)}$} & $\mathcal{U}_\mathrm{b}(30,40)$\newline$\mathcal{U}_\mathrm{c}(100,120)$\newline$\mathcal{N}_\mathrm{Prot, GP}(96,6)$ & 36.1186\newline113.4634\newline97.6017 &            -382.7 &                     0.0 &        -341.9 \\
\noalign{\smallskip}
\hline
\end{tabular}
\tablefoot{The model used for the final fit is the {2P$_\text{(36\,d-ecc, 113\,d-circ)}$ + dSHO-GP$_\text{(96\,d)}$} model highlighted in boldface. {Wide, unconstrained priors are denoted as ``blind.''}}
\end{table*}}

\begin{figure*}[!ht]
\centering
\includegraphics[scale=0.8]{legend_rv_plots.pdf}\\
\includegraphics[width=0.45\textwidth]{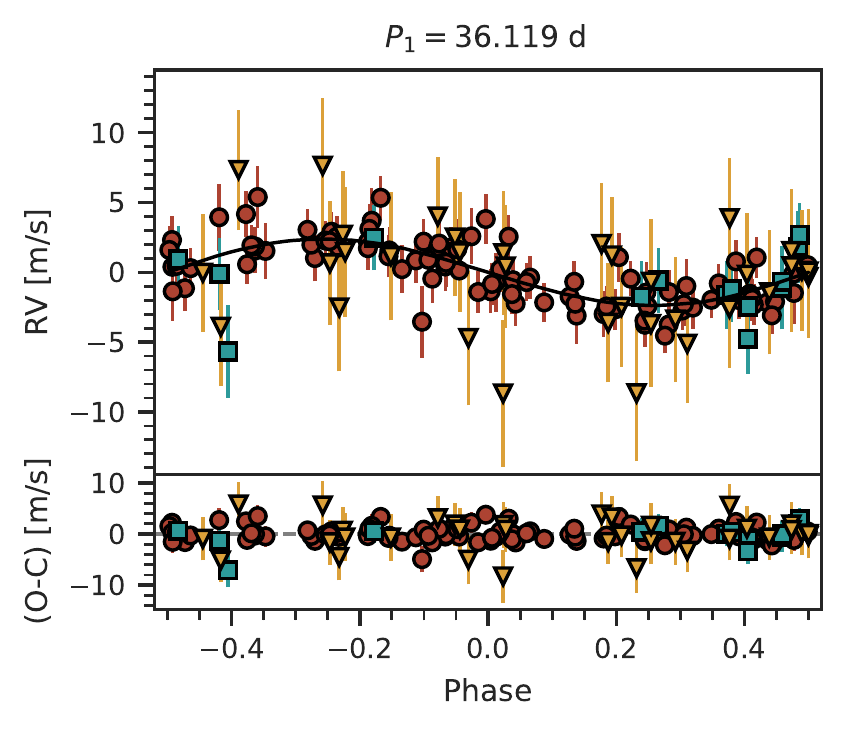}
\includegraphics[width=0.45\textwidth]{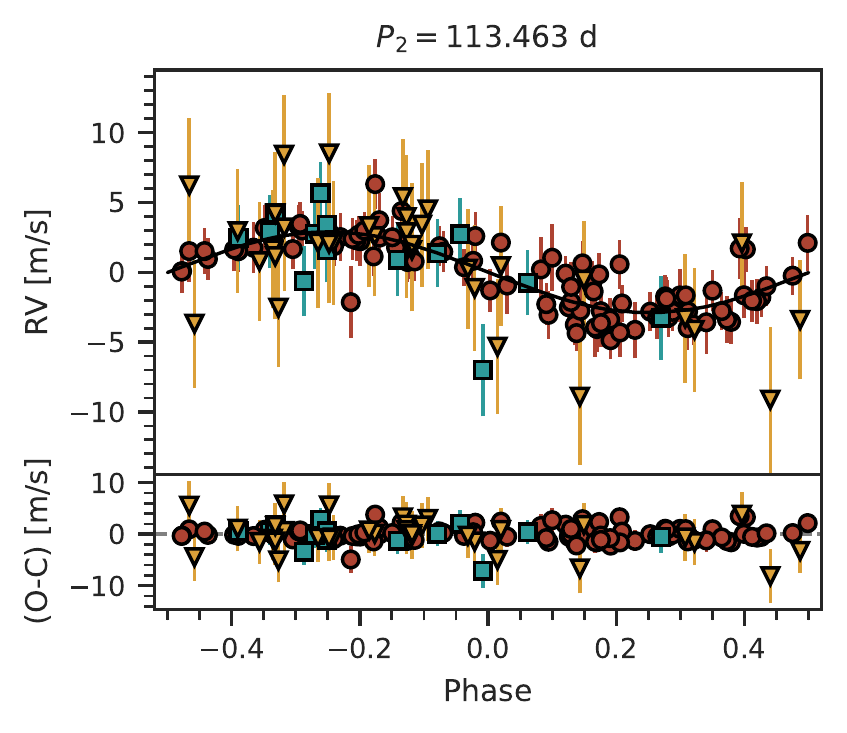}
\caption{Phased RV model for the \hoststar{} planets from the final {2P$_\text{(36\,d-ecc, 113\,d-circ)}$ + dSHO-GP$_\text{(96\,d)}$} model. The black lines show the median of \num{10000} samples from the posterior. The residuals after subtracting the median models are shown in the lower panel.} 
\label{fig:fit_rv_phased}
\end{figure*}

We found two {potentially planetary} signals in the periodogram analysis: {an isolated} signal with a period of 36.1\,d and {a second signal close to the} stellar rotation with a period of 113.1\,d. Following this, we conducted a model comparison based on the Bayesian evidence to find the model that explains the RV data the best. Our base assumption was that the RV data can be fully explained by stellar activity (= ``0P-model''). For this model, we used GP modeling as in \autoref{subsec:rotation} to account for the quasi-periodic nature of such signals. {To ensure consistency with the photometry, we applied the dSHO kernel in our analysis.} Therefore, we used individual GP amplitudes for each instrument to consider the different scatter {in} the HIRES, HARPS, and CARMENES data. Furthermore, we made use of the determined stellar rotation period and adopted a normal prior centered on our most precise measurement from the dSHO kernel of $P_\text{rot}\approx$ {96\,d,} with {three times the} uncertainty of {the photometric fit} ($\pm 6$\,d). We compared this activity-only model with two more complex models that assume the presence of further signals in the data. The first one additionally includes the \Pbday{} signal using a Keplerian component with a uniform period prior between 30\,d and 40\,d (hereafter, the ``1P-model''). We set {$t_0$, the time of inferior conjunction}, between BJD = 2457433 and 2457450 to avoid a split posterior due to the aliasing. Further, we re-parametrized the eccentricity and the argument of periastron by $\mathcal{S}_1 = \sqrt{e}\sin\omega$ and $\mathcal{S}_2 = \sqrt{e}\cos\omega$ {to allow for a uniform sampling} between \num{-1} and \num{1} \citep{Eastman2013}. In another fit, we {additionally considered the \Pcday{} signal} by adding a second Keplerian component to the model. For this, we used {a period prior between 100\,d and 120\,d} and {$t_0$} between BJD = {2457433 and 2457550} (hereafter, the ``2P-model''). For both 1P and 2P models, we tested circular as well as eccentric models for the Keplerian signals. For all instruments, we set individual uninformed uniform priors for the RV offset and the jitter. A {summary} of the used priors can be found in \autoref{tab:hnlib_priors_details}.

We found that the models including the \Pbday{} signal are generally significantly \citep[i.e., $\Delta\ln\mathcal{Z}>5$;][]{Trotta2008}
better than the model considering only the stellar activity (see \autoref{tab:modelcomp}). {Likewise, the models including two Keplerian signals are mostly favored against the 1P-models, except for the model in which both planetary orbits are assumed to be eccentric. The highest evidence corresponds to the two planetary model, one eccentric orbit at 36\,d and another circular orbit at 113\,d, plus the dSHO kernel of $P_\text{rot}$ at 96\,d (2P$_\text{(36\,d-ecc, 113\,d-circ)}$ + dSHO-GP$_\text{(96\,d)}$), which we chose as our final model for the further analysis  (shown in Figs.~\ref{fig:fit_rv_full} and~\ref{fig:fit_rv_phased}). The inner planet orbits in a nearly circular orbit of eccentricity $0.079^{0.090}_{0.055}$. The posteriors from the fit for the planetary parameters and the GP can be found in \autoref{tab:pl_posterior} and \autoref{fig:cornerplot_final_fit}, and the instrumental posteriors in \autoref{tab:instr_posterior}. However, regardless of which model is chosen, all determined period and semi-amplitude combinations of the \Pbday{} signal, {which is the focus of our analysis}, agree within their uncertainties (see \autoref{fig:modelcomparison}). 

\begin{table}
    \caption{Posteriors of the planetary parameters and the GP component from the final model.}
    \label{tab:pl_posterior}
    \centering
    {\setlength{\extrarowheight}{4.5pt}
        \begin{tabular}{cccc}
            \hline \hline
            Parameter                                   & Planet b & Planet [c]                   & Units                  \\
            \hline
            \noalign{\smallskip}
            \multicolumn{4}{c}{\textit{Planetary posteriors } }                                                                       \\
            \noalign{\smallskip}
            $P$                                & $\num{36.116}^{+\num{0.027}}_{-\num{0.029}}$ &  $\num{113.46}^{+\num{0.19}}_{-\num{0.20}}$ & d                      \\
            $t_{0}$                            & $\num{2457436.3}^{+\num{1.4}}_{-\num{1.4}}$  & $\num{2457492.8}^{+\num{3.6}}_{-\num{3.3}}$  & d                      \\
            $\sqrt{e}\cos \omega$     & $\num{0.03}^{+\num{0.22}}_{-\num{0.23}}$  &  \dots   & \dots                  \\
            $\sqrt{e}\sin \omega$     & $\num{-0.11}^{+\num{0.23}}_{-\num{0.21}}$ &  \dots   & \dots                  \\
            $K$                                & $\num{2.59}^{+\num{0.34}}_{-\num{0.34}}$  &  $\num{2.92}^{+\num{0.54}}_{-\num{0.56}}$   & $\mathrm{m\,s^{-1}}$   \\
            \noalign{\smallskip}
            \multicolumn{4}{c}{\textit{Derived planetary parameters} }                                                               \\
            \noalign{\smallskip}
            $M\sin i$                           & $\num{5.46}^{+\num{0.75}}_{-\num{0.75}}$ & $\num{9.7}^{+\num{1.9}}_{-\num{1.9}}$      & M$_\oplus$             \\
            $e$                                & $\num{0.079}^{+\num{0.090}}_{-\num{0.055}}$ & \dots   & \dots \\
            $a$                                 & $\num{0.1417}^{+\num{0.0023}}_{-\num{0.0023}}$ & $\num{0.3040}^{+\num{0.0048}}_{-\num{0.0051}}$ & \si{\astronomicalunit} \\
            $S$                                         & $\num{0.503}^{+\num{0.018}}_{-\num{0.016}}$  & $\num{0.1094}^{+\num{0.0039}}_{-\num{0.0035}}$   & S$_\oplus$             \\
            $T_\text{eq}(1-A)^{-1/4}$& $\num{234.4}^{+\num{5.5}}_{-\num{5.3}}$ & $\num{160.1}^{+\num{3.7}}_{-\num{3.7}}$       & \si{\kelvin}           \\
            \noalign{\smallskip}
            \multicolumn{4}{c}{\textit{GP posteriors} }                                                                           \\
            \noalign{\smallskip}
            $\sigma_\text{GP, CARMENES}$                & \multicolumn{2}{c}{$\num{3.47}^{+\num{1.04}}_{-\num{0.67}}$}       & $\mathrm{m\,s^{-1}}$   \\
            $\sigma_\text{GP, HIRES}$                   & \multicolumn{2}{c}{$\num{6.88}^{+\num{1.73}}_{-\num{2.00}}$}       & $\mathrm{m\,s^{-1}}$   \\
            $\sigma_\text{GP, HARPS}$                   & \multicolumn{2}{c}{$\num{2.35}^{+\num{1.51}}_{-\num{1.03}}$}       & $\mathrm{m\,s^{-1}}$   \\
            $f_\text{GP}$                               & \multicolumn{2}{c}{$\num{0.33}^{+\num{0.38}}_{-\num{0.25}}$}       & \dots                  \\
            $Q_{0, \text{GP}}$                          & \multicolumn{2}{c}{$\num{0.22}^{+\num{0.37}}_{-\num{0.10}}$}       & \dots                  \\
            $dQ_\text{GP}$                              & \multicolumn{2}{c}{$\num{122.00}^{+\num{23938}}_{-\num{114}}$}     & \dots                  \\
            $P_\text{rot, GP}$                          & \multicolumn{2}{c}{$\num{96\pm2}$}     & d                      \\
            \noalign{\smallskip}
            \hline
        \end{tabular}}
    \tablefoot{{Error bars denote the $68\%$ posterior {confidence} intervals.}
    }
\end{table}

\subsection{Investigation of the \Pcday{} signal}
\label{subsec:113d}

{Our findings about the \Pcday{} signal up to this point can be summarized as follows: We found a signal in the RVs that is close to, but not exactly at, the period of the stellar rotation period, which we determined consistently from photometry and spectroscopic activity indicators. Furthermore, the stellar rotation period of \Prot{} and the period in question at \Pc{} are not related by aliasing. In fact, we see evidence for a periodic signal in the RVs that coincides with the stellar rotation period that is distinct in period from the yearly aliases of the \Pcday{} signal. Our analysis of the signal's sBGLS periodogram and the model comparison show that it is very well described by a static Keplerian model.}

\subsection{Wavelength dependence}
\begin{figure}
\centering
\includegraphics{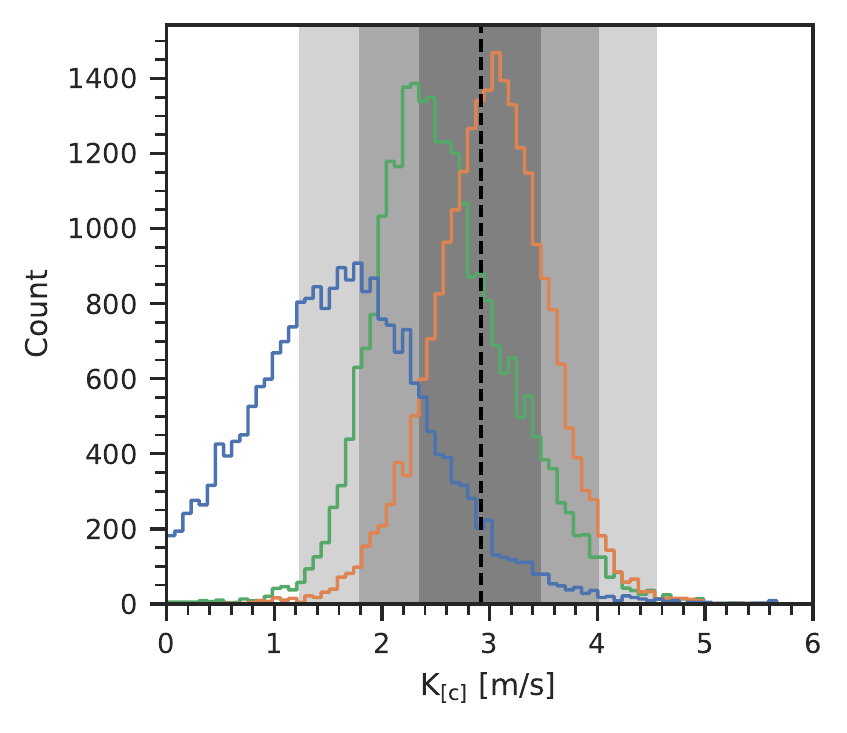}
\caption{{RV} semi-amplitudes of the \Pcday{} signal (\planetc{}) as a function of wavelength for three different wavelength chunks of the whole CARMENES VIS wavelength domain. {Blue:} 5612--6491\,\AA, {green:} 6448--7613\,\AA, {orange:} 7576--9203\,\AA. The black dashed line indicates the best fit from \autoref{subsec:rvmodelling}, while the grey shades highlight the 1, 2, and 3\,$\sigma$ region around it.}  
\label{fig:rv_amplitude_comparison}
\end{figure}

{In the presence of strong stellar activity, we would expect a wavelength dependence of the RVs. A good first indicator for activity is the CRX, which expresses the RV-$\log\lambda$-correlation \citep{Zechmeister2018, Tal-Or2018, Jeffers2022}. We used the CRX calculated from the CARMENES visual channel data to investigate the presence of stellar activity in \hoststar{}. However, the GLS periodogram of the CRX data does not show any peaks near the period in question (see \autoref{fig:gls_crx}). Further, both in the original data and after removing the \Pbday{} signal, there is no significant correlation between the RVs and the CRX.}

{Another method to investigate a potential wavelength dependence of a RV signal is to make use of the RVs computed for each \'echelle order \citep{Bauer2020}. To do this, we divided the CARMENES RVs into three roughly equal wavelength ranges and performed a fit to each of them to determine the semi-amplitude of the \Pcday{} signal. For the fit, we took the results from the best fit in \autoref{subsec:rvmodelling} as normal priors for the \Pbday{} signal and the dSHO-GP kernel, as well as the period of the \Pcday{} signal. The prior of the semi-amplitude of the \Pcday{} signal was set uniform between 0\,m\,s$^{-1}$ and 10\,m\,s$^{-1}$. The results for the posterior distribution from those fits are depicted in \autoref{fig:rv_amplitude_comparison}. There are no significant discrepancies. The increase in amplitude and, at the same time, improvement in precision with increasing wavelength can be explained by the lower RV information content in the bluer parts of the spectrum of the M4.0\,V host star \citep[e.g.,][and references therein]{Reiners2020}.}

\section{Discussion}
\label{sec:discussion}

\subsection{HN Lib b}
\label{subsec:hnlib_b}

\begin{figure}
\centering
\includegraphics[width=\columnwidth]{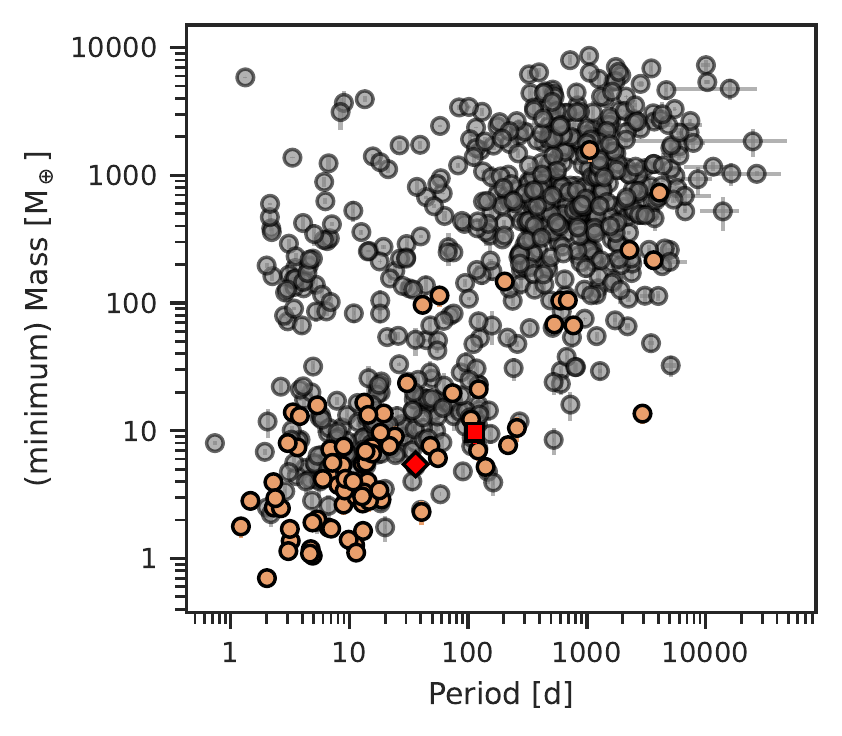}
\caption{Known exoplanets discovered with the RV method. Planets orbiting M dwarfs are marked with orange and the rest with grey circles. \planetb{} is highlighted by a red diamond {and the planet candidate \planetc{} by a red square}. For planets only detected by RVs, the minimum mass is depicted, while for planetary system with known inclination (e.g., from transits), the true mass is given. Based on the NASA Exoplanet Archive.} 
\label{fig:period-mass}
\end{figure}

\begin{figure}
\centering
\includegraphics[width=\columnwidth]{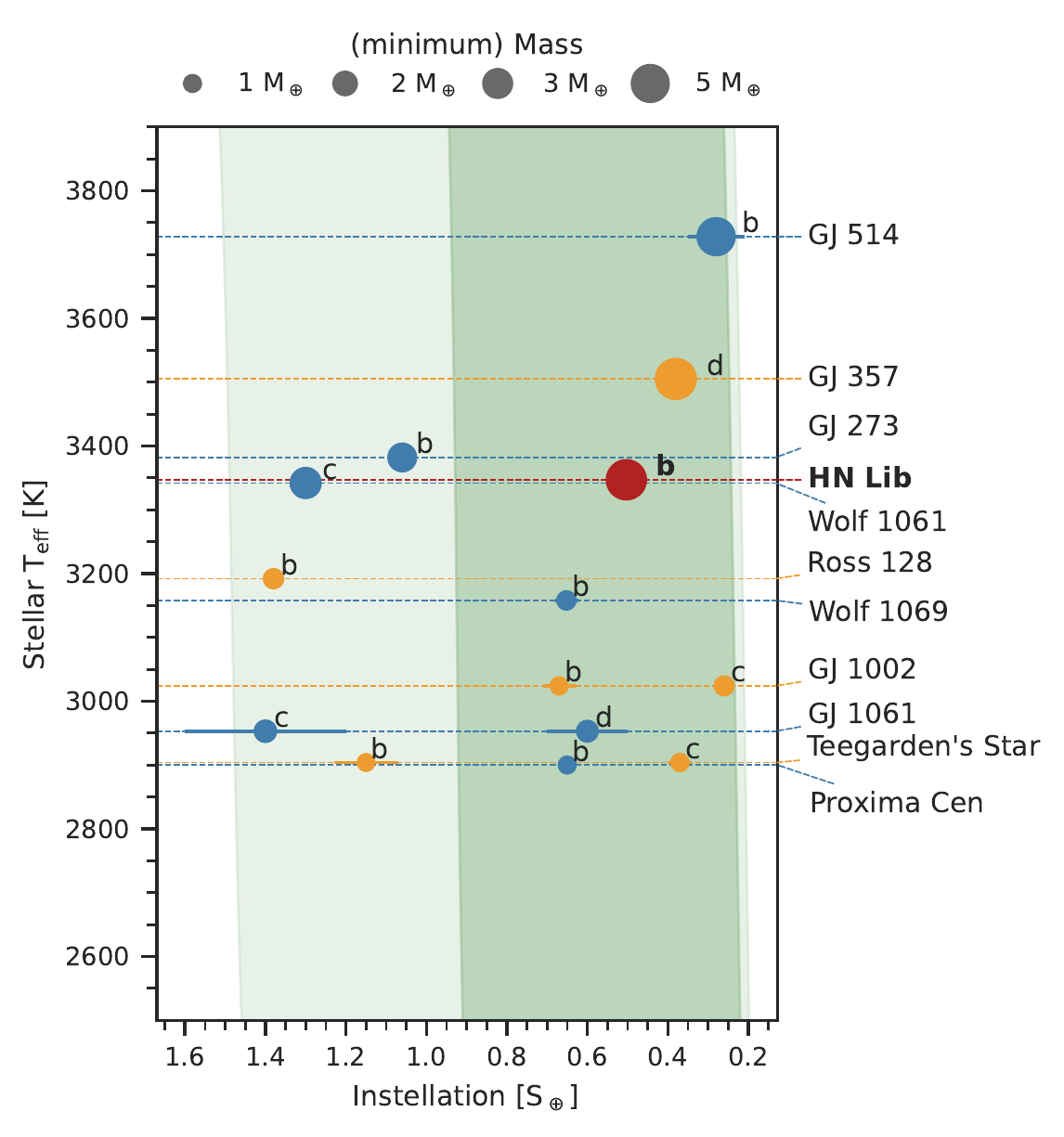}
\caption{Comparison to other small (i.e., $M\sin{i} \leq 10 \text{M}_\oplus$) {potentially} habitable planets orbiting M-dwarf stars at less than 10\,pc from the Sun. \planetb{} is highlighted in red, while the other planets are iterated between orange and blue for better distinguishability. The light green and dark green shaded areas denote the optimistic and conservative HZs of \citet{Kopparapu2013, Kopparapu2014}, respectively. {The planet list was} taken from the habitable planets catalog of the Planetary Habitability Laboratory at the University of Puerto Rico and merged with the composite planetary parameters table of the NASA Exoplanet Archive.} 
\label{fig:habplanets}
\end{figure}

The monitoring of the M dwarf \hoststar{} during our observing campaign within the CARMENES GTO program yielded a sub-Neptunian-mass planet discovery, namely \planetb{}. The minimum mass of the planet is $M_{\rm b}\sin i = $ \Mbfit{}, and it orbits its host star in a nearly circular orbit at a separation of $a_{\rm b}=0.1417\pm0.0023\,\text{au}$ with an orbital period of $P_\text{b} = $ \Pbfit{}.

\planetb{}'s signal is unrelated to the stellar rotation period {that we determined to be $P_\text{rot} =$ \Prot{}} and has no significant counterparts in the activity indicators. Furthermore, the stability and the coherence of the {36-d} signal indicate a long-lived behavior, stable over the time span of our observations and thus providing strong arguments in favor of the planetary nature of \planetb{}. As it can be seen in \autoref{fig:period-mass}, the planet adds to the sample of small, sub-Neptunian mass planets, which make up the bulk of planets detected to orbit M dwarfs on short orbits and that are predicted to be the most abundant planet type around M dwarfs \citep{Dressing2015, Sabotta2021, Pinamonti2022}.

\cite{Kopparapu2013, Kopparapu2014} calculated conservative estimations of the inner HZ around stars with stellar effective temperatures in the range 2600–-7200\,K for planetary masses between 0.1 and 5\,M$_\oplus$. {Taking into account} the runaway greenhouse effect for 5\,M$_\oplus$ {planets such as \planetb{}, the inner edge of \hoststar{}'s HZ is accordingly placed at 0.101\,au}. With an instellation of $S \approx 0.50\,S_\oplus$ and a semi-major axis of 0.142\,au, \planetb{} thus resides in the conservative HZ of \hoststar. At a distance of only 6.25\,pc, \hoststar{} is one of the closest systems to Earth with a planet in the conservative HZ.

In the two extreme cases {of the Bond albedo}, A = 0.0 and A = 0.65, the $T_{\rm eq}$ for \planetb{} {falls in} the range {of} 243--174\,K for a non-reflecting planet and for a highly reflecting planet, respectively. In \autoref{fig:habplanets} we compare the position of \planetb{} with other known habitable zone sub-Neptune-type and super-Earth planets with (minimum) mass measurements around M dwarfs. However, as for all planets in this regime, the actual habitability depends on many factors. The most decisive is probably the nature of its atmosphere \citep{Kaltenegger2017}. 

Unfortunately, \hoststar{} has not been observed by NASA's Transiting Exoplanet Survey Satellite (\textit{TESS}), nor  are any such observations planned for future \textit{TESS} sectors. Given the ambiguity of the mass-radius relation for small planets \citep[e.g.,][]{Fulton2017, VanEylen2018, Petigura2020, Cloutier2020}, we thus only can hypothesize about its size and composition. Following the recent empirical analysis of small, transiting planets with mass measurements by \cite{Luque2022}, there are three different plausible compositions  given the mass of \planetb{}: a rocky planet with an Earth-like density, a water-rich planet, and a gaseous planet with an extended hydrogen atmosphere. Applying the theoretical mass-radius relationships by \cite{Zeng2019} to the different possibilities results in radii of $\sim 1.6$\,R$_\oplus$ for an Earth-like planet, $\sim 2.0$\,R$_\oplus$ for a water-rich planet, and $\sim 2.4$\,R$_\oplus$ for 1$\%$ of hydrogen added to an Earth-like core (assuming an equilibrium temperature of $300$\,K in all three cases). The resulting Earth similarity index \citep{Schulze-Makuch2011} for a rocky planet would be 0.60, comparable to TRAPPIST-1\,g \citep{Gillon2017, Grimm2018, Ducrot2020, Agol2021} or the recently discovered GJ~1002\,b \citep{SuarezMascareno2023}. For a full transit, the depth would be in the range of $\sim$1600--5400\,ppm taking into account the different possible compositions. However, the probability that our planet transits is low (0.98$^{+1.02}_{-0.94}$\%).

We explored the high energy environment of \planetb{} to test for the habitability conditions on the planet. There are no positive X-ray detections of \hoststar{} in the literature \citep{Wood1994, Schmitt1995}, but \citet{Stelzer2013} calculated an upper limit of $\log L_{\rm X}<26.80$. As an alternative approach, we used the rotation period of \Prot{} to calculate the expected X-ray emission, based on the relations of \citet{Wright2011}, resulting in $\log L_{\rm X}\sim 26.9$, in approximate agreement with the {aforementioned} upper limit. We then used the relationships in \citet{Sanz-Forcada2011} to calculate the extreme ultraviolet (EUV) luminosity ($L_{\rm EUV}$) corresponding to this X-ray flux, $\log L_{\rm EUV}=27.9$, and the expected (upper limit) mass loss rate in an energy-limited approach, $1.8 \times 10^9$ g\,s$^{-1}$ or 0.0097\,M$_\oplus$\,Gyr$^{-1}$. This mass loss rate is too low to pose any problem for the present stability of the planet's atmosphere, which is consistent with the planet's location in the {potentially HZ} of \hoststar{}. {The} stellar age based on the X-ray luminosity is quite uncertain \citep[$\sim$7~Gyr,][]{Sanz-Forcada2011}, but it is not consistent with values as low as 0.8\,Gyr, which would otherwise imply X-ray activity approaching   saturation, $\log L_{\rm X}\sim 28.4$, and a shorter rotation period.

\subsection{HN Lib [c]}
{The RV observations show another significant signal at a period of $\sim$\Pc{}, close to the stellar rotation period of $96 \pm 2$\,d. Our model comparison shows that the signal is reasonably well described by a circular Keplerian orbit and, following the sBGLS analysis, seems to be reasonably stable over the whole period of our observations,  spanning 21\,years. When analyzing its wavelength dependency, we found no significant changes in the amplitude of the signal that would hint at its origins coming from stellar activity. However, given the proximity to the stellar rotation period, we could not fully rule out this scenario (especially due to the presence of differential rotation) and thus report it as a planet candidate. \planetc{} has a minimum mass of \Mcfit{} and orbits \hoststar{} at a separation of $0.3040\,\pm0.0051\,\text{au}$. In contrast to \planetb{}, the orbit lies outside the HZ of \hoststar{}. As for planet b, predictions about its composition are ambiguous because it resides in the same mass regime. However, due to the orbit beyond the ice line of \hoststar{}, an ice-rich planet is very likely \citep{Burn2021}. In combination with the also comparatively massive inner companion, the planetary system of \hoststar{} is therefore very interesting with regard to the formation and evolution of systems with planets in the HZ.}

\section{Summary}
\label{sec:summary}
The newly discovered planet \planetb{} orbits a nearby M4.0\,V star and has a minimum mass of $M_{\rm b}\sin i = $ \Mbfit{}. Given the ambiguity of the radius in this mass range, this planet could be either rocky, a water-rich, or gaseous. The orbital period of the planet, $P_\text{b} = $ \Pbfit{}, corresponds to a separation to its host of $a_{\rm b}=0.1417\pm0.0023\,\text{au}$ centered in the conservative HZ of \hoststar{} ($S \approx 0.50\,S_\oplus$). 
\planetb{} is one of the closest planets orbiting its host in the HZ as seen from Earth and, assuming that it has a rocky composition, its Earth similarity index would be 0.60.

Additionally, our RV measurements show another significant signal with a period of $P_\text{[c]}= $ \Pcfit{}, which is close to the stellar rotation period of $P_\text{rot}=$ \Prot{} that we determined from photometric measurements -- and which was additionally validated with spectroscopic activity indicators. Since we can neither confirm nor exclude that the signal is an imprint of stellar activity,  although we do know is stable over the time span of our observations, we report it to be a planet candidate with a minimum mass of \Mcfit{}.


\begin{acknowledgements}

We wish to thank the anonymous referee for helpful comments and suggestions, which helped to improve the manuscript. CARMENES is an instrument at the Centro Astron\'omico Hispano en Andaluc\'ia (CAHA) at Calar Alto (Almer\'{\i}a, Spain), operated jointly by the Junta de Andaluc\'ia and the Instituto de Astrof\'isica de Andaluc\'ia (CSIC).

CARMENES was funded by the German Max-Planck- Gesellschaft (MPG), the Spanish Consejo Superior de Investigaciones Cient\'ificas (CSIC), the European Union through FEDER/ERF funds, and the members of the CARMENES Consortium (Max-Planck-Institut f\"ur Astronomie, Instituto de Astrof\'isica de Andaluc\'ia, Landessternwarte Ko\"onigstuhl, Institut de Ci\` encies de l'Espai, Insitut f\"ur Astrophysik G\"ottingen, Universidad, Complutense de Madrid, Th\"uringer Landessternwarte Tautenburg, Instituto de Astrof\'isica de Canarias, Hamburger Sternwarte, Centro de Astrobiolog\'ia and Centro Astron\'omico Hispano-Alem\'an), with additional contributions by the Spanish Ministry of Economy, the state of Baden-W\"uttemberg and Niedersachsen, the Klaus Tschira Foundation (KTS), the Deutsche Forschungsgemeinschaft (DFG) via the Research Unit FOR2544 ``Blue Planets around Red Stars'', and by the Junta de Andaluc\'ia. 
This work was based on data from the CARMENES data archive at CAB (CSIC-INTA). Data were partly collected with the 90\,cm telescope at the Sierra Nevada Observatory (OSN) operated by the Instituto de Astrof\'\i fica de Andaluc\'\i a (IAA-CSIC). We acknowledge the telescope operators from Observatori Astron\'omic del Montsec, Sierra Nevada Observatory, and CAHA. 
We acknowledge financial support from the Agencia Estatal de Investigaci\'on 10.13039/501100011033 of the Ministerio de Ciencia e Innovaci\'on and the ERDF ``A way of making Europe'' through projects 
  PID2019-109522GB-C5[1,2,3,4]  
  PGC2018-098153-B-C3[1,3]              
and the Centre of Excellence ``Severo Ochoa'' and ``Mar\'ia de Maeztu'' awards to the Instituto de Astrof\'isica de Canarias (CEX2019-000920-S), Instituto de Astrof\'isica de Andaluc\'ia (SEV-2017-0709), and Centro de Astrobiolog\'ia (MDM-2017-0737), the Generalitat de Catalunya/CERCA programme,
and the Israel Science Foundation (grant No. 1404/22).

This research has made use of the NASA Exoplanet Archive, which is operated by the California Institute of Technology, under contract with the National Aeronautics and Space Administration under the Exoplanet Exploration Program.

\end{acknowledgements}

\bibliographystyle{aa} 
\bibliography{HNlib.bib} 
\clearpage


\begin{appendix} 
\onecolumn
\section{Additional figures}
\begin{figure*}[!ht]
        \centering
        \includegraphics[width=\textwidth]{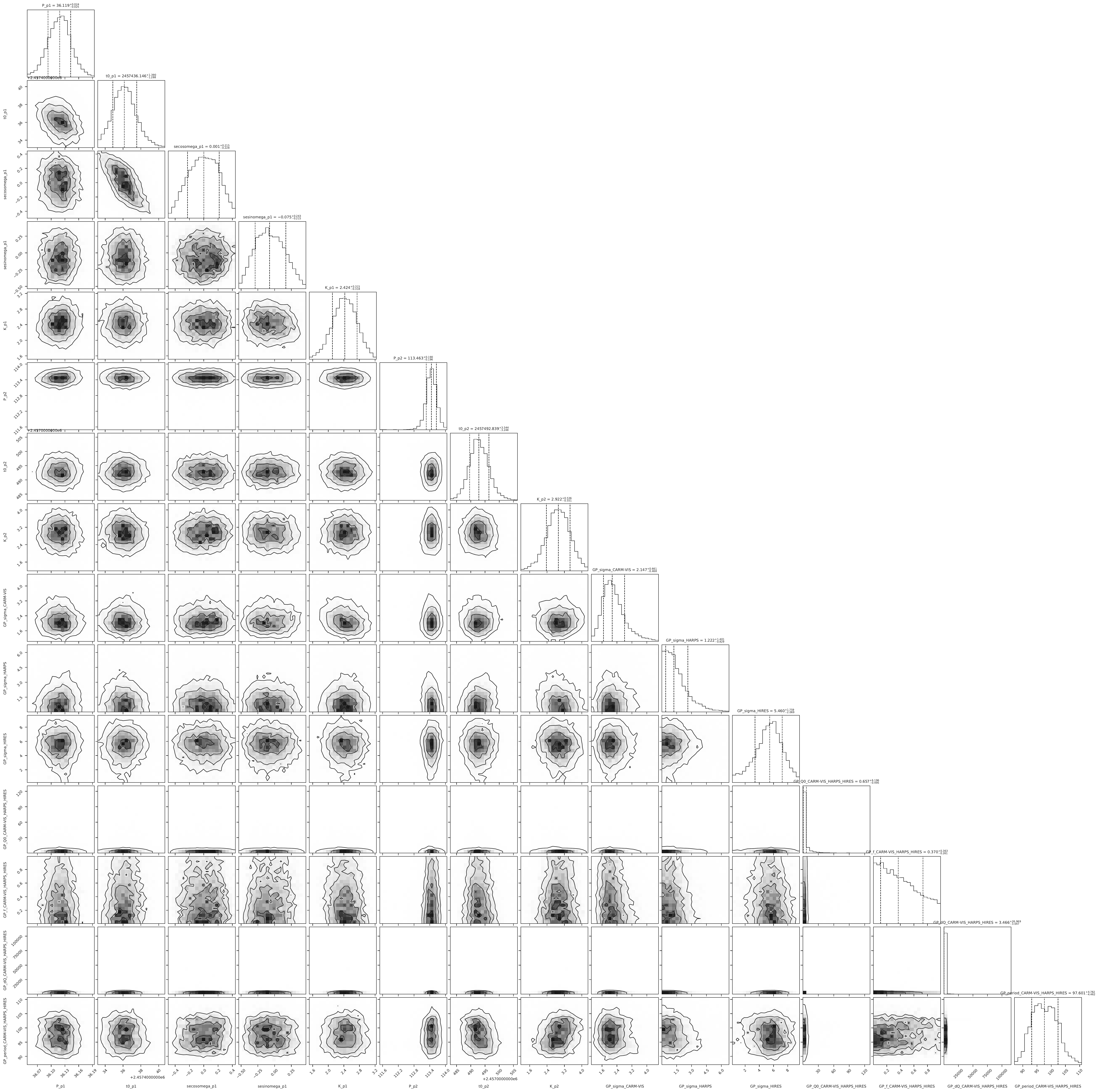}
        \caption{Posterior distributions of the fitted planetary and GP parameters of the \hoststar{} system as obtained from the final 1P$_\text{(36 d-ecc)}$ + dSHO-GP$_\text{(113 d)}$  fit. The vertical dashed lines indicate the 16, 50, and 84\%~quantiles that were used to define the optimal values and their associated 1$\sigma$ uncertainty. 
        }
        \label{fig:cornerplot_final_fit}
\end{figure*}

\begin{figure}[!ht]
        \centering
        \includegraphics{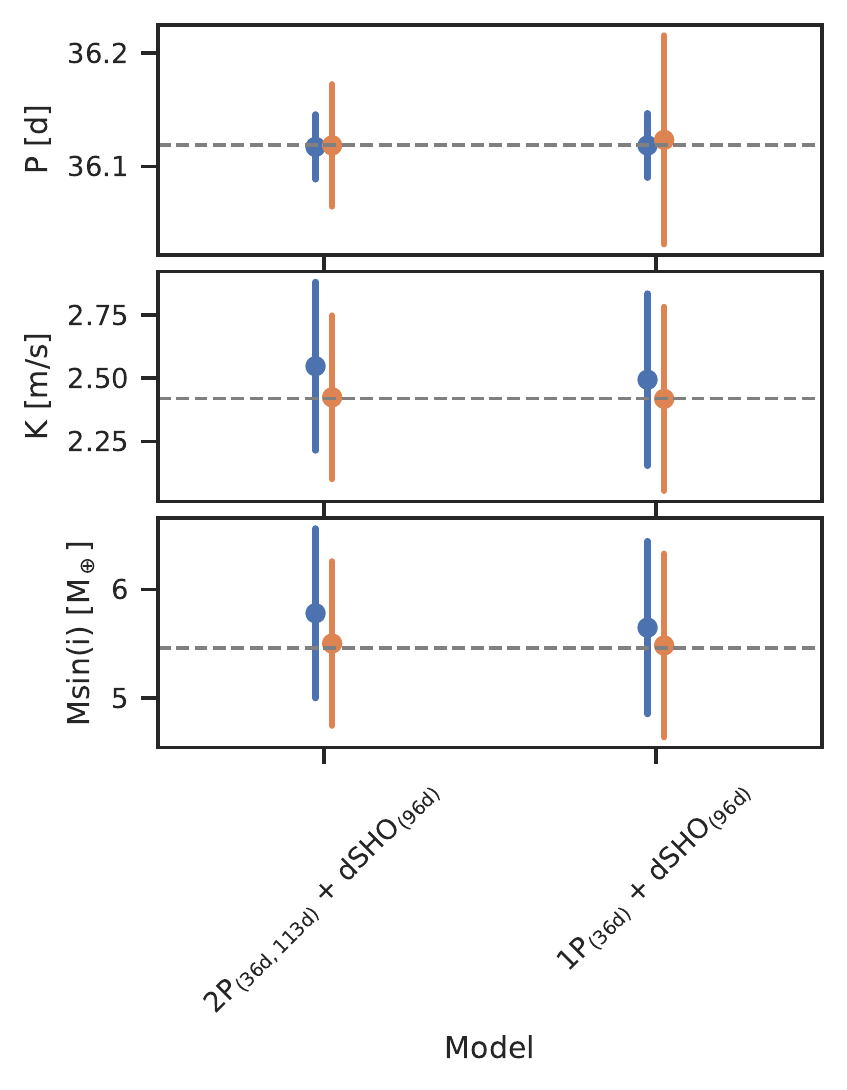}
        \caption{Posterior distributions of the period, semi-amplitude, and determined mass of the \Pbday{} planet for the different models. The blue markers depict the circular models and the orange markers the eccentric models. 
        }
        \label{fig:modelcomparison}
\end{figure}

\begin{figure}[!ht]
        \centering
        \includegraphics{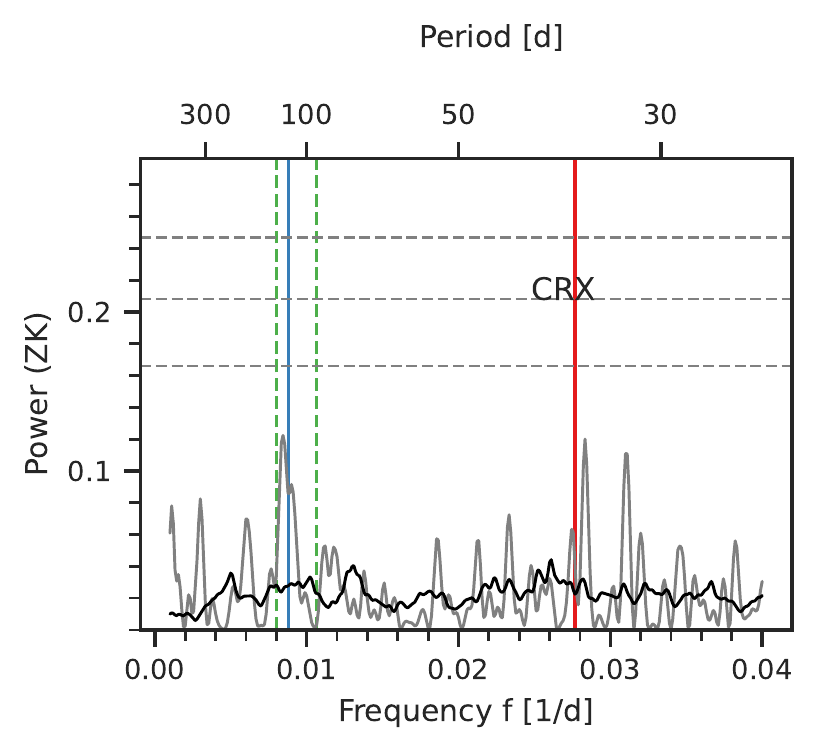}
        \caption{GLS periodogram of the CRX activity indicator.}
        \label{fig:gls_crx}
\end{figure}

\begin{figure*}[!ht]
        \centering
        \includegraphics{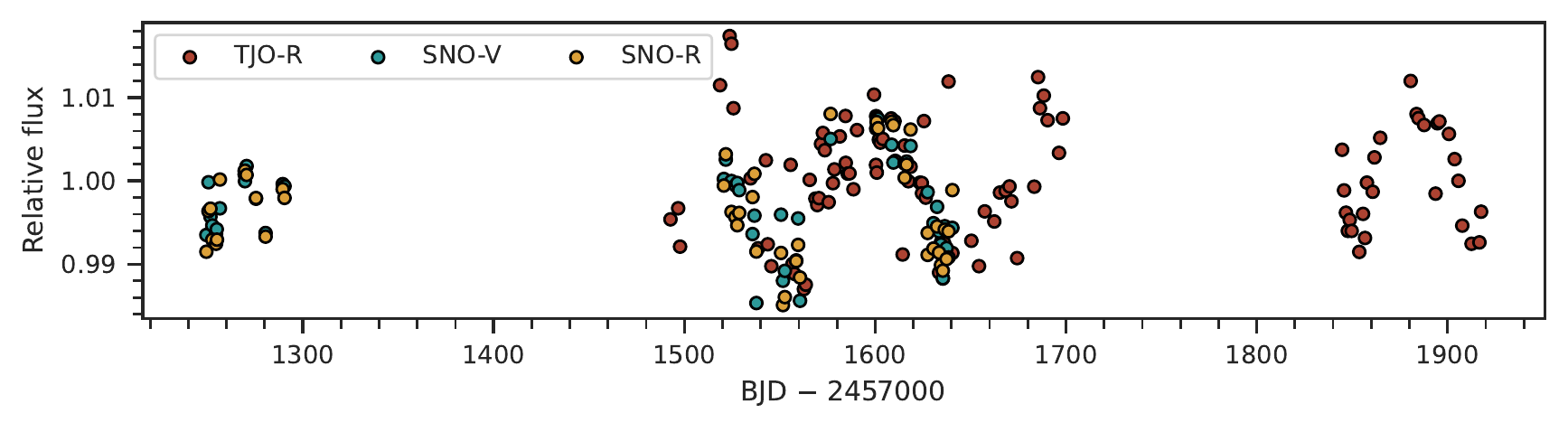}
        \caption{Comparison of the TJO and OSN photometry after the CAFM detrending. 
        }
        \label{fig:comparison_TJO-OSN}
\end{figure*}

\clearpage

\section{Tables}
{\setlength{\extrarowheight}{4.5pt}

\begin{table*}[!ht]
    \centering
    \caption{Priors used to determine the photometric rotation period of \hoststar.}
    \label{tab:priors_phot}
    \begin{tabular}{l c c p{12cm}}
        \hline
        \hline
        \noalign{\smallskip}
        Parameter                        & Prior                               & Unit            & Description                                                                \\
        \noalign{\smallskip}
        \hline
        \noalign{\smallskip}
        \multicolumn{4}{c}{\textit{dSHO-GP parameters}}                                                                                                                            \\
        \noalign{\smallskip}
        $\sigma_\text{GP}$     & $\mathcal{J}(1,10^6 )$            & ppm  & Standard deviation of the GP, individual for each instrument but shared between the MEarth telescopes              \\
        $f_\text{GP}$                    & $\mathcal{U}(0.0, 1.0)$             & \dots           & Fractional amplitude of the secondary mode compared to the primary mode, shared between all instruments    \\
        $Q_{0, \text{GP}}$               & $\mathcal{J}(0.1, 10^{10})$       & \dots           & Quality factor of the secondary oscillation, shared between all instruments                                \\
        $dQ_\text{GP}$                   & $\mathcal{J}(0.1, 10^{10})$       & \dots           & Difference in quality factor between the primary and secondary oscillation, shared between all instruments \\
        $P_\text{rot, GP}$               & $\mathcal{U}(10,200)$          & d               & Period of the GP kernel, shared between all instruments                                                    \\
        \noalign{\smallskip}
        \multicolumn{4}{c}{\textit{QP-GP parameters}}                                                                                                                            \\
        \noalign{\smallskip}
        $\sigma_\text{GP}$     & $\mathcal{J}(1,10^6 )$            & ppm  & Standard deviation of the GP, individual for each instrument but shared between the MEarth telescopes              \\
        $\alpha_{\text{GP}}$               & $\mathcal{J}(10^{-10}, 1)$       & \dots           & Correlation timescale of the exponential part of the kernel, shared between all instruments                                \\
        $\Gamma_\text{GP}$                   & $\mathcal{J}(0, 10)$       & \dots          & Harmonic complexity of the sine part of the kernel, shared between all instruments \\
        $P_\text{rot, GP}$               & $\mathcal{U}(110,120)$          & d               & Period of the GP kernel, shared between all instruments                                                    \\
        \noalign{\smallskip}
        \multicolumn{4}{c}{\textit{Instrument parameters}}                                                                                                                    \\
        \noalign{\smallskip}

        mflux                            & $\mathcal{N}(0, 0.2)$        & \dots & Median relative flux                                \\
        $\sigma$                         & $\mathcal{J}(1, 10^6)$       & ppm & A jitter added in quadrature to each instrument                  \\
        \noalign{\smallskip}
        \hline
    \end{tabular}
    \tablefoot{The prior labels $\mathcal{U}$, $\mathcal{J}$, and $\mathcal{N}$ represent uniform, log-uniform, and normal distributions, respectively.}
\end{table*}

\begin{table*}[!ht]
    \centering
    \caption{Priors used for the RV fit of \hoststar.}
    \label{tab:hnlib_priors_details}
    \begin{tabular}{l c c p{10cm}}
        \hline
        \hline
        \noalign{\smallskip}
        Parameter                        & Prior                               & Unit            & Description                                                                \\
        \noalign{\smallskip}
        \hline
        \noalign{\smallskip}
        \multicolumn{4}{c}{\textit{Planet parameters} }                                                                                                                   \\
        \noalign{\smallskip}

        $P_\text{b}$                              & $\mathcal{U}(30, 40)$           & d               & Period                                                                     \\
        $K_\text{b}$                              & $\mathcal{U}(0, 10)$            & m\,$\rm s^{-1}$ & RV semi-amplitude                                                          \\
        $t_\text{0, b}$ (BJD)                      & $\mathcal{U}(2457433, 2457450)$ & d               & Time of periastron passage                                                 \\
        $\sqrt{e_\text{b}}\sin \omega_\text{b}$ & $\mathcal{U}(-1.0, 1.0)$            & \dots           & {Parameterization} for $e$ and $\omega$                                   \\
        $\sqrt{e_\text{b}}\cos \omega_\text{b}$ & $\mathcal{U}(-1.0, 1.0)$            & \dots           & {Parameterization} for $e$ and $\omega$                                   \\
        $P_\text{c}$                              & $\mathcal{U}(100, 120)$           & d               & Period                                                                     \\
        $K_\text{c}$                              & $\mathcal{U}(0.0, 10.0)$            & m\,$\rm s^{-1}$ & RV semi-amplitude                                                          \\   
        $t_\text{0, c}$ (BJD)                      & $\mathcal{U}(2457433, 2457550)$ & d               & Time of periastron passage                                                 \\        
        $\sqrt{e_\text{c}}\sin \omega_\text{c}$ & fixed (0)           & \dots           & {Parameterization} for $e$ and $\omega$                                   \\
        $\sqrt{e_\text{c}}\cos \omega_\text{c}$ & fixed (0)           & \dots           & {Parameterization} for $e$ and $\omega$                                   \\

        \noalign{\smallskip}
        \multicolumn{4}{c}{\textit{GP parameters}}                                                                                                                            \\
        \noalign{\smallskip}
        $\sigma_\text{GP}$     & $\mathcal{U}(0.0, 10)$            & m\,$\rm s^{-1}$ & Standard deviation of the GP, individual for each instrument               \\
        $f_\text{GP}$                    & $\mathcal{U}(0.0, 1.0)$             & \dots           & Fractional amplitude of the secondary mode compared to the primary mode, shared between all instruments    \\
        $Q_{0, \text{GP}}$               & $\mathcal{J}(0.1, 10^6)$       & \dots           & Quality factor of the secondary oscillation, shared between all instruments                                \\
        $dQ_\text{GP}$                   & $\mathcal{J}(0.1, 10^6)$       & \dots           & Difference in quality factor between the primary and secondary oscillation, shared between all instruments \\
        $P_\text{rot, GP}$               & $\mathcal{N}(96, 6)$          & d               & Period of the GP kernel, shared between all instruments                                                    \\

        \noalign{\smallskip}
        \multicolumn{4}{c}{\textit{Instrument parameters}}                                                                                                                    \\
        \noalign{\smallskip}

        $\gamma$                            & $\mathcal{U}(-10, 10)$          & m\,$\rm s^{-1}$ & RV zero point for CARMENES, HIRES, and HARPS                                \\
        $\sigma$                         & $\mathcal{U}(0.0, 10)$            & m\,$\rm s^{-1}$ & A jitter added in quadrature to CARMENES, HIRES, and HARPS                  \\
        \noalign{\smallskip}
        \hline
    \end{tabular}
    \tablefoot{The prior labels $\mathcal{U}$, $\mathcal{J}$, and $\mathcal{N}$ represent uniform, log-uniform, and normal distributions, respectively.}
\end{table*}

    \begin{table*}[!ht]
    \caption{Instrumental posteriors from the final model.}
    \label{tab:instr_posterior}
    \begin{tabular}[t]{lcl}
    \hline
    \hline
    Parameter & Posterior\tablefootmark{(a)} & Units \\
    \hline
        \noalign{\smallskip}
     \multicolumn{3}{c}{CARMENES} \\
        $\gamma$           & $\num{0.76}^{+\num{0.3}}_{-\num{0.3}}$             & \si{\meter\per\second}    \\
        $\sigma$        & $\num{0.91}^{+\num{0.31}}_{-\num{0.3}}$            & \si{\meter\per\second}                       \\
        \noalign{\smallskip}
        \multicolumn{3}{c}{HARPS} \\
        $\gamma$           & $\num{-0.8}^{+\num{1.2}}_{-\num{1.3}}$             & \si{\meter\per\second}    \\
        $\sigma$        & $\num{2.04}^{+\num{0.93}}_{-\num{0.7}}$            & \si{\meter\per\second}                       \\
        \noalign{\smallskip}
        \multicolumn{3}{c}{HIRES} \\
        $\gamma$           & $\num{-3.6}^{+\num{1.4}}_{-\num{1.5}}$             & \si{\meter\per\second}    \\
        $\sigma$        & $\num{3.4}^{+\num{1.8}}_{-\num{1.6}}$              & \si{\meter\per\second}                       \\
    \hline
    \end{tabular}  
\tablefoot{\tablefoottext{a}{Error bars denote the $68\%$ posterior {confidence} intervals.}}    
    \end{table*}
}

\clearpage

\begin{landscape}
\section{Data}
{\scriptsize                                                
\begin{longtable}[!ht]{S[table-format=7.5]S[table-format=-2.2(2)]S[table-format=1.5(5)]S[table-format=1.5(5)]S[table-format=1.5(5)]S[table-format=1.5(5)]S[table-format=1.5(5)]S[table-format=1.5(5)]S[table-format=1.5(5)]}
\caption{RVs and spectral activity indicators of \hoststar{} obtained with CARMENES} \label{tab:hnlib_rv_act_data} \\
\toprule
{BJD} & {RV [ms$^{-1}$]} & {Ca~{\sc ii} IRT b [$\AA$]} & {He $\lambda 10830\,\AA$ [$\AA$]} & {Pa $\beta$ [$\AA$]} & {TiO $\lambda 7048\,\AA$ [$\AA$]} & {TiO $\lambda 8428\,\AA$ [$\AA$]} & {TiO $\lambda 8858\,\AA$ [$\AA$]} & {VO $\lambda 7940\,\AA$ [$\AA$]} \\
\midrule
\endfirsthead
\toprule
{BJD} & {RV [ms$^{-1}$]} & {Ca~{\sc ii} IRT b [$\AA$]} & {He $\lambda 10830\,\AA$ [$\AA$]} & {Pa $\beta$ [$\AA$]} & {TiO $\lambda 7048\,\AA$ [$\AA$]} & {TiO $\lambda 8428\,\AA$ [$\AA$]} & {TiO $\lambda 8858\,\AA$ [$\AA$]} & {VO $\lambda 7940\,\AA$ [$\AA$]} \\
\midrule
\endhead
\midrule
\multicolumn{9}{r}{Continued on next page} \\
\midrule
\endfoot
\bottomrule
\endlastfoot
2457414.76773 & -3.9\pm1.2 & 0.5877\pm0.0022 & {\dots} & {\dots} & 0.5006\pm0.0011 & 0.8081\pm0.0020 & 0.8000\pm0.0018 & 0.8810\pm0.0029 \\
2457415.75313 & -4.5\pm1.2 & 0.5803\pm0.0017 & {\dots} & {\dots} & 0.49804\pm0.00083 & 0.8038\pm0.0016 & 0.8021\pm0.0014 & 0.8773\pm0.0023 \\
2457427.73064 & 1.1\pm1.0 & 0.5797\pm0.0013 & 0.00512\pm0.00097 & -0.0089\pm0.0034 & 0.50267\pm0.00064 & 0.8074\pm0.0012 & 0.8005\pm0.0011 & 0.8787\pm0.0018 \\
2457466.70322 & 1.7\pm1.4 & 0.6062\pm0.0019 & 0.01077\pm0.00087 & -0.0101\pm0.0036 & 0.50030\pm0.00090 & 0.8103\pm0.0017 & 0.8082\pm0.0015 & 0.8798\pm0.0024 \\
2457468.54591 & -3.5\pm2.4 & 0.5935\pm0.0023 & 0.0030\pm0.0010 & -0.0074\pm0.0031 & 0.5028\pm0.0011 & 0.8018\pm0.0021 & 0.8091\pm0.0019 & 0.8805\pm0.0029 \\
2457473.66410 & -0.4\pm1.3 & 0.5798\pm0.0017 & 0.01123\pm0.00078 & -0.0080\pm0.0038 & 0.50721\pm0.00079 & 0.8115\pm0.0015 & 0.8120\pm0.0014 & 0.8797\pm0.0022 \\
2457490.58135 & 4.5\pm1.4 & 0.5831\pm0.0025 & 0.0167\pm0.0011 & -0.0065\pm0.0017 & 0.4994\pm0.0013 & 0.8118\pm0.0023 & 0.8082\pm0.0020 & 0.8808\pm0.0033 \\
2457503.53506 & 2.0\pm1.4 & 0.5879\pm0.0015 & 0.01082\pm0.00072 & -0.0067\pm0.0024 & 0.50099\pm0.00075 & 0.8058\pm0.0014 & 0.8067\pm0.0013 & 0.8862\pm0.0020 \\
2457509.54415 & 4.1\pm1.3 & 0.5883\pm0.0020 & {\dots} & {\dots} & 0.50194\pm0.00099 & 0.8060\pm0.0018 & 0.8056\pm0.0016 & 0.8786\pm0.0027 \\
2457538.45706 & 7.6\pm1.3 & 0.5920\pm0.0024 & 0.01452\pm0.00092 & -0.00358\pm0.00066 & 0.4967\pm0.0012 & 0.8016\pm0.0021 & 0.8165\pm0.0020 & 0.8835\pm0.0032 \\
2457540.43448 & 3.1\pm1.3 & 0.5979\pm0.0011 & 0.01729\pm0.00100 & -0.00431\pm0.00075 & 0.49894\pm0.00054 & 0.80193\pm0.00094 & 0.82471\pm0.00091 & 0.8796\pm0.0014 \\
2457553.41742 & -0.1\pm2.0 & 0.5866\pm0.0031 & 0.00837\pm0.00089 & 0.00585\pm0.00096 & 0.4940\pm0.0016 & 0.8006\pm0.0027 & 0.8275\pm0.0025 & 0.8815\pm0.0041 \\
2457588.34937 & {\dots} & 0.578\pm0.014 & {\dots} & {\dots} & {\dots} & {\dots} & {\dots} & {\dots} \\
2457762.76898 & -6.4\pm1.3 & 0.5843\pm0.0019 & 0.0018\pm0.0011 & -0.0098\pm0.0035 & 0.50511\pm0.00096 & 0.8128\pm0.0018 & 0.8012\pm0.0016 & 0.8785\pm0.0026 \\
2457793.70241 & 6.1\pm1.4 & 0.5860\pm0.0019 & 0.00689\pm0.00074 & -0.0108\pm0.0034 & 0.50084\pm0.00098 & 0.8062\pm0.0018 & 0.8025\pm0.0016 & 0.8781\pm0.0027 \\
2457798.67671 & 3.0\pm1.1 & 0.5862\pm0.0017 & 0.0057\pm0.0020 & -0.0109\pm0.0039 & 0.50284\pm0.00080 & 0.8106\pm0.0016 & 0.8030\pm0.0014 & 0.8889\pm0.0023 \\
2457799.70019 & 4.3\pm1.2 & 0.5880\pm0.0021 & 0.0060\pm0.0022 & -0.0113\pm0.0035 & 0.5013\pm0.0010 & 0.8095\pm0.0019 & 0.7987\pm0.0017 & 0.8767\pm0.0028 \\
2457824.63108 & 6.2\pm1.1 & 0.5939\pm0.0013 & 0.01269\pm0.00088 & -0.0065\pm0.0035 & 0.49843\pm0.00064 & 0.8087\pm0.0012 & 0.8057\pm0.0011 & 0.8785\pm0.0018 \\
2457830.59583 & 3.1\pm1.1 & 0.5913\pm0.0015 & 0.0184\pm0.0020 & -0.0103\pm0.0036 & 0.50262\pm0.00076 & 0.8113\pm0.0014 & 0.8061\pm0.0013 & 0.8778\pm0.0021 \\
2457833.60155 & -0.6\pm1.3 & 0.5932\pm0.0011 & 0.01280\pm0.00097 & -0.0061\pm0.0034 & 0.50020\pm0.00055 & 0.8103\pm0.0010 & 0.81441\pm0.00095 & 0.8810\pm0.0016 \\
2457850.55617 & {\dots} & 0.5884\pm0.0013 & {\dots} & {\dots} & 0.50429\pm0.00064 & 0.8119\pm0.0012 & 0.8035\pm0.0011 & 0.8797\pm0.0018 \\
2457851.61608 & -1.3\pm1.4 & 0.5877\pm0.0016 & 0.01315\pm0.00087 & -0.0101\pm0.0023 & 0.50086\pm0.00080 & 0.8091\pm0.0015 & 0.8031\pm0.0013 & 0.8784\pm0.0022 \\
2457852.49788 & -4.2\pm1.4 & 0.5851\pm0.0014 & 0.01385\pm0.00075 & -0.0102\pm0.0029 & 0.50020\pm0.00069 & 0.8086\pm0.0013 & 0.8024\pm0.0011 & 0.8789\pm0.0019 \\
2457892.55145 & 2.9\pm1.3 & 0.5891\pm0.0015 & 0.0086\pm0.0010 & -0.00446\pm0.00084 & 0.49900\pm0.00071 & 0.8020\pm0.0013 & 0.8138\pm0.0012 & 0.8795\pm0.0019 \\
2458140.73748 & 4.7\pm1.0 & 0.5792\pm0.0012 & 0.00996\pm0.00090 & -0.0138\pm0.0056 & 0.50499\pm0.00062 & 0.8079\pm0.0012 & 0.8022\pm0.0010 & 0.8797\pm0.0017 \\
2458199.62582 & -6.6\pm1.8 & 0.5804\pm0.0029 & 0.00912\pm0.00093 & -0.0094\pm0.0026 & 0.5044\pm0.0015 & 0.8080\pm0.0027 & 0.8030\pm0.0023 & 0.8772\pm0.0040 \\
2458205.58323 & -6.1\pm1.1 & 0.5840\pm0.0012 & 0.01023\pm0.00070 & -0.0091\pm0.0036 & 0.50640\pm0.00059 & 0.8124\pm0.0011 & 0.8087\pm0.0010 & 0.8816\pm0.0017 \\
2458245.56524 & 1.2\pm1.6 & 0.5850\pm0.0014 & 0.02224\pm0.00082 & -0.00516\pm0.00082 & 0.50151\pm0.00068 & 0.8055\pm0.0013 & 0.8115\pm0.0012 & 0.8771\pm0.0019 \\
2458264.46262 & 3.4\pm1.5 & 0.5786\pm0.0015 & 0.01397\pm0.00072 & -0.00378\pm0.00067 & 0.50192\pm0.00073 & 0.8054\pm0.0013 & 0.8149\pm0.0012 & 0.8799\pm0.0020 \\
2458271.46979 & 0.9\pm1.1 & 0.5842\pm0.0012 & 0.01214\pm0.00083 & -0.00346\pm0.00066 & 0.50175\pm0.00058 & 0.8024\pm0.0011 & 0.8270\pm0.0010 & 0.8783\pm0.0016 \\
2458273.38241 & -0.5\pm1.4 & 0.5856\pm0.0013 & 0.01345\pm0.00080 & -0.00256\pm0.00070 & 0.49840\pm0.00063 & 0.8012\pm0.0012 & 0.8361\pm0.0011 & 0.8834\pm0.0017 \\
2458289.35756 & 5.0\pm1.4 & 0.5910\pm0.0013 & 0.01071\pm0.00081 & 0.00202\pm0.00072 & 0.49720\pm0.00060 & 0.8002\pm0.0011 & 0.8395\pm0.0011 & 0.8801\pm0.0017 \\
2458290.45492 & 2.2\pm1.8 & 0.5883\pm0.0015 & 0.00043\pm0.00075 & 0.00819\pm0.00079 & 0.49184\pm0.00072 & 0.7987\pm0.0013 & 0.8446\pm0.0013 & 0.8795\pm0.0020 \\
2458296.43821 & 3.0\pm2.2 & 0.5883\pm0.0012 & -0.01891\pm0.00069 & 0.00652\pm0.00073 & 0.49456\pm0.00057 & 0.8026\pm0.0011 & 0.8443\pm0.0011 & 0.8788\pm0.0016 \\
2458297.38299 & 0.3\pm1.2 & 0.5877\pm0.0012 & -0.01002\pm0.00063 & 0.00493\pm0.00074 & 0.49671\pm0.00061 & 0.8048\pm0.0011 & 0.8375\pm0.0011 & 0.8787\pm0.0017 \\
2458301.37002 & -1.4\pm1.0 & 0.5889\pm0.0017 & -0.0291\pm0.0026 & 0.0004\pm0.0011 & 0.49944\pm0.00085 & 0.8074\pm0.0015 & 0.8178\pm0.0014 & 0.8849\pm0.0023 \\
2458302.43405 & -3.0\pm1.2 & 0.5888\pm0.0016 & -0.08227\pm0.00082 & 0.00171\pm0.00086 & 0.49943\pm0.00077 & 0.8045\pm0.0014 & 0.8215\pm0.0014 & 0.8792\pm0.0021 \\
2458303.43433 & -2.4\pm1.9 & 0.5811\pm0.0032 & -0.1492\pm0.0043 & 0.0021\pm0.0021 & 0.4909\pm0.0015 & 0.8042\pm0.0028 & 0.8352\pm0.0026 & 0.8803\pm0.0041 \\
2458310.36421 & -0.6\pm1.5 & 0.5878\pm0.0015 & -0.03997\pm0.00067 & 0.00016\pm0.00076 & 0.49559\pm0.00076 & 0.8101\pm0.0014 & 0.8254\pm0.0013 & 0.8776\pm0.0021 \\
2458329.36043 & -0.7\pm1.4 & 0.5844\pm0.0016 & -0.08658\pm0.00080 & -0.00066\pm0.00082 & 0.49766\pm0.00079 & 0.8107\pm0.0014 & 0.8226\pm0.0013 & 0.8782\pm0.0021 \\
2458332.34518 & 0.7\pm1.4 & 0.5866\pm0.0014 & -0.06505\pm0.00080 & -0.00375\pm0.00072 & 0.49500\pm0.00068 & 0.8098\pm0.0013 & 0.8296\pm0.0012 & 0.8746\pm0.0018 \\
2458480.76686 & 4.6\pm1.3 & 0.5872\pm0.0014 & 0.00584\pm0.00093 & -0.0089\pm0.0040 & 0.50598\pm0.00068 & 0.8114\pm0.0013 & 0.8044\pm0.0012 & 0.8787\pm0.0019 \\
2458485.74859 & 1.6\pm1.0 & 0.5746\pm0.0014 & 0.00778\pm0.00086 & -0.0154\pm0.0064 & 0.50268\pm0.00071 & 0.8121\pm0.0013 & 0.8013\pm0.0012 & 0.8820\pm0.0020 \\
2458486.74758 & -0.0\pm1.1 & 0.5866\pm0.0013 & 0.00931\pm0.00075 & -0.0146\pm0.0043 & 0.50300\pm0.00063 & 0.8092\pm0.0012 & 0.7997\pm0.0011 & 0.8818\pm0.0017 \\
2458490.74074 & -1.0\pm1.2 & 0.5858\pm0.0013 & 0.01213\pm0.00078 & -0.0228\pm0.0083 & 0.50157\pm0.00064 & 0.8073\pm0.0012 & 0.8012\pm0.0011 & 0.8853\pm0.0018 \\
2458493.76022 & -2.3\pm1.1 & 0.5791\pm0.0015 & 0.01102\pm0.00084 & -0.0099\pm0.0039 & 0.50142\pm0.00074 & 0.8075\pm0.0014 & 0.8037\pm0.0012 & 0.8809\pm0.0021 \\
2458494.75537 & 0.3\pm1.7 & 0.5705\pm0.0031 & -0.0293\pm0.0013 & -0.0089\pm0.0018 & 0.5021\pm0.0015 & 0.8032\pm0.0029 & 0.8050\pm0.0025 & 0.8924\pm0.0042 \\
2458497.72975 & -0.9\pm1.0 & 0.5742\pm0.0013 & 0.0041\pm0.0015 & -0.0107\pm0.0039 & 0.50059\pm0.00064 & 0.8084\pm0.0012 & 0.8057\pm0.0011 & 0.8784\pm0.0018 \\
2458498.71549 & 2.1\pm1.1 & 0.5811\pm0.0017 & 0.0059\pm0.0010 & -0.0123\pm0.0038 & 0.50198\pm0.00084 & 0.8076\pm0.0016 & 0.8003\pm0.0014 & 0.8796\pm0.0023 \\
2458499.74934 & -1.1\pm1.1 & 0.5794\pm0.0011 & 0.0062\pm0.0014 & -0.0097\pm0.0036 & 0.50415\pm0.00057 & 0.8080\pm0.0011 & 0.80561\pm0.00096 & 0.8808\pm0.0016 \\
2458509.75153 & 2.16\pm0.97 & 0.5807\pm0.0013 & 0.00335\pm0.00095 & -0.0096\pm0.0033 & 0.50337\pm0.00062 & 0.8096\pm0.0012 & 0.8007\pm0.0011 & 0.8802\pm0.0018 \\
2458527.73930 & -1.65\pm0.84 & 0.5787\pm0.0012 & 0.0033\pm0.0011 & -0.0113\pm0.0037 & 0.50068\pm0.00063 & 0.8101\pm0.0012 & 0.8047\pm0.0011 & 0.8835\pm0.0018 \\
2458528.68192 & -3.69\pm0.85 & 0.5762\pm0.0012 & 0.0073\pm0.0012 & -0.0117\pm0.0039 & 0.50476\pm0.00060 & 0.8100\pm0.0012 & 0.8012\pm0.0010 & 0.8797\pm0.0017 \\
2458529.66779 & -5.88\pm0.88 & 0.5801\pm0.0013 & 0.00629\pm0.00097 & -0.0108\pm0.0039 & 0.50139\pm0.00063 & 0.8109\pm0.0012 & 0.8017\pm0.0011 & 0.8783\pm0.0017 \\
2458532.69776 & -2.3\pm1.1 & 0.5770\pm0.0014 & 0.0038\pm0.0013 & -0.0102\pm0.0033 & 0.49978\pm0.00074 & 0.8094\pm0.0014 & 0.8070\pm0.0012 & 0.8815\pm0.0021 \\
2458533.66697 & -0.7\pm1.2 & 0.5755\pm0.0017 & 0.0009\pm0.0015 & -0.0095\pm0.0035 & 0.50282\pm0.00085 & 0.8090\pm0.0016 & 0.8108\pm0.0014 & 0.8793\pm0.0024 \\
2458534.67050 & -3.8\pm1.1 & 0.5801\pm0.0013 & 0.0095\pm0.0012 & -0.0081\pm0.0036 & 0.50284\pm0.00067 & 0.8116\pm0.0013 & 0.8049\pm0.0011 & 0.8798\pm0.0019 \\
2458535.68757 & -4.61\pm0.97 & 0.5799\pm0.0012 & 0.00586\pm0.00097 & -0.0131\pm0.0038 & 0.50151\pm0.00059 & 0.8075\pm0.0012 & 0.8026\pm0.0010 & 0.8780\pm0.0017 \\
2458537.71021 & -0.94\pm0.98 & 0.5847\pm0.0013 & 0.00667\pm0.00092 & -0.0088\pm0.0032 & 0.50525\pm0.00063 & 0.8100\pm0.0012 & 0.7994\pm0.0011 & 0.8806\pm0.0018 \\
2458542.69476 & 0.8\pm1.1 & 0.5828\pm0.0013 & 0.00894\pm0.00083 & -0.0127\pm0.0038 & 0.50311\pm0.00064 & 0.8091\pm0.0012 & 0.7994\pm0.0011 & 0.8815\pm0.0018 \\
2458543.65836 & {\dots} & 0.5808\pm0.0012 & 0.0064\pm0.0010 & -0.0118\pm0.0038 & 0.50380\pm0.00061 & 0.8097\pm0.0012 & 0.8073\pm0.0010 & 0.8819\pm0.0017 \\
2458543.65942 & {\dots} & 0.5808\pm0.0012 & 0.0064\pm0.0010 & -0.0118\pm0.0038 & 0.50380\pm0.00061 & 0.8097\pm0.0012 & 0.8073\pm0.0010 & 0.8819\pm0.0017 \\
2458545.65316 & 2.4\pm1.2 & 0.5768\pm0.0016 & 0.00633\pm0.00082 & -0.0105\pm0.0035 & 0.50094\pm0.00079 & 0.8118\pm0.0015 & 0.8029\pm0.0013 & 0.8794\pm0.0022 \\
2458546.65916 & 1.8\pm1.0 & 0.5828\pm0.0019 & 0.01198\pm0.00095 & -0.0095\pm0.0034 & 0.49993\pm0.00093 & 0.8071\pm0.0018 & 0.8005\pm0.0015 & 0.8782\pm0.0026 \\
2458553.72719 & 2.1\pm1.1 & 0.5848\pm0.0012 & 0.0091\pm0.0012 & -0.0050\pm0.0042 & 0.49946\pm0.00059 & 0.8083\pm0.0011 & 0.81035\pm0.00098 & 0.8776\pm0.0016 \\
2458560.73491 & -1.4\pm1.2 & 0.5760\pm0.0014 & 0.0078\pm0.0011 & -0.0132\pm0.0082 & 0.50532\pm0.00071 & 0.8105\pm0.0014 & 0.8085\pm0.0012 & 0.8807\pm0.0019 \\
2458589.57015 & 3.2\pm1.5 & 0.5796\pm0.0022 & 0.0087\pm0.0010 & -0.0063\pm0.0020 & 0.5022\pm0.0011 & 0.8099\pm0.0021 & 0.8020\pm0.0018 & 0.8769\pm0.0030 \\
2458603.48060 & -0.5\pm1.2 & 0.5833\pm0.0012 & 0.00929\pm0.00081 & -0.0075\pm0.0017 & 0.50249\pm0.00058 & 0.8057\pm0.0011 & 0.8130\pm0.0010 & 0.8808\pm0.0016 \\
2458604.52155 & -0.06\pm0.96 & 0.5816\pm0.0012 & 0.00773\pm0.00082 & -0.0047\pm0.0011 & 0.50048\pm0.00061 & 0.8052\pm0.0011 & 0.8125\pm0.0010 & 0.8805\pm0.0017 \\
2458605.57208 & 0.49\pm0.94 & 0.5805\pm0.0012 & 0.00772\pm0.00096 & -0.00434\pm0.00094 & 0.50332\pm0.00060 & 0.8044\pm0.0011 & 0.8088\pm0.0010 & 0.8834\pm0.0017 \\
2458610.51896 & 1.8\pm1.2 & 0.5840\pm0.0013 & 0.00995\pm0.00078 & -0.00457\pm0.00074 & 0.50280\pm0.00065 & 0.8071\pm0.0012 & 0.8068\pm0.0011 & 0.8820\pm0.0018 \\
2458614.48521 & 1.5\pm1.0 & 0.5825\pm0.0013 & 0.00741\pm0.00090 & -0.00349\pm0.00068 & 0.50194\pm0.00063 & 0.8095\pm0.0012 & 0.8123\pm0.0011 & 0.8814\pm0.0017 \\
2458619.54597 & 3.0\pm1.1 & 0.5833\pm0.0012 & 0.00843\pm0.00097 & -0.00444\pm0.00078 & 0.50248\pm0.00060 & 0.8051\pm0.0011 & 0.81619\pm0.00100 & 0.8814\pm0.0016 \\
2458646.41563 & -2.9\pm1.9 & 0.5791\pm0.0021 & 0.0047\pm0.0011 & 0.00443\pm0.00074 & 0.4976\pm0.0011 & 0.8008\pm0.0019 & 0.8270\pm0.0017 & 0.8850\pm0.0028 \\
2458647.42417 & -1.3\pm1.2 & 0.5879\pm0.0013 & 0.0158\pm0.0019 & 0.0012\pm0.0012 & 0.49784\pm0.00063 & 0.8011\pm0.0012 & 0.8347\pm0.0011 & 0.8760\pm0.0017 \\
2458657.40840 & 1.0\pm1.3 & 0.5873\pm0.0013 & 0.00810\pm0.00086 & 0.00646\pm0.00068 & 0.49473\pm0.00062 & 0.8033\pm0.0011 & 0.8401\pm0.0011 & 0.8782\pm0.0017 \\
2458661.42292 & 1.8\pm1.7 & 0.5871\pm0.0021 & -0.0457\pm0.0010 & -0.0003\pm0.0100 & 0.4938\pm0.0011 & 0.7993\pm0.0019 & 0.8390\pm0.0018 & 0.8814\pm0.0029 \\
2458662.44232 & 1.5\pm1.9 & 0.5823\pm0.0028 & -0.1002\pm0.0013 & 0.0059\pm0.0010 & 0.4938\pm0.0014 & 0.7963\pm0.0025 & 0.8416\pm0.0023 & 0.8705\pm0.0036 \\
2458681.37492 & 0.9\pm1.0 & 0.5829\pm0.0014 & -0.06744\pm0.00091 & -0.00273\pm0.00088 & 0.49371\pm0.00066 & 0.8067\pm0.0013 & 0.8385\pm0.0012 & 0.8788\pm0.0018 \\
2458691.36881 & 3.6\pm1.2 & 0.5835\pm0.0018 & -0.0728\pm0.0010 & -0.00606\pm0.00084 & 0.49101\pm0.00089 & 0.8086\pm0.0017 & 0.8428\pm0.0016 & 0.8778\pm0.0024 \\
2458865.74522 & 4.7\pm2.2 & 0.5760\pm0.0039 & -0.0128\pm0.0018 & -0.0071\pm0.0028 & 0.5024\pm0.0020 & 0.8049\pm0.0037 & 0.8093\pm0.0031 & 0.8712\pm0.0053 \\
2458877.74747 & -3.1\pm1.5 & 0.5745\pm0.0013 & -0.0004\pm0.0014 & -0.0120\pm0.0036 & 0.49933\pm0.00066 & 0.8072\pm0.0013 & 0.8048\pm0.0011 & 0.8761\pm0.0018 \\
2458881.73734 & {\dots} & 0.589\pm0.018 & {\dots} & {\dots} & 0.5114\pm0.0087 & 0.797\pm0.016 & 0.808\pm0.012 & 0.868\pm0.019 \\
2458885.74379 & -4.7\pm1.2 & 0.5720\pm0.0014 & 0.00005\pm0.00091 & -0.0102\pm0.0035 & 0.50364\pm0.00068 & 0.8112\pm0.0013 & 0.8034\pm0.0012 & 0.8794\pm0.0019 \\
2458889.72751 & -8.0\pm1.3 & 0.5763\pm0.0014 & -0.0047\pm0.0014 & -0.0093\pm0.0036 & 0.50339\pm0.00070 & 0.8111\pm0.0013 & 0.8048\pm0.0012 & 0.8811\pm0.0020 \\
2458895.74796 & -5.9\pm1.0 & 0.5709\pm0.0013 & 0.00168\pm0.00097 & -0.0148\pm0.0057 & 0.50604\pm0.00063 & 0.8108\pm0.0012 & 0.8071\pm0.0011 & 0.8817\pm0.0018 \\
2458903.67784 & -0.4\pm1.1 & 0.5796\pm0.0012 & 0.00446\pm0.00075 & -0.0128\pm0.0038 & 0.50339\pm0.00060 & 0.8090\pm0.0012 & 0.8037\pm0.0010 & 0.8813\pm0.0017 \\
2458913.64500 & 1.1\pm1.1 & 0.5830\pm0.0014 & 0.00979\pm0.00073 & -0.0092\pm0.0036 & 0.50364\pm0.00071 & 0.8120\pm0.0014 & 0.8052\pm0.0012 & 0.8831\pm0.0020 \\
2458917.69630 & 1.4\pm1.4 & 0.5868\pm0.0021 & 0.00156\pm0.00096 & -0.0046\pm0.0033 & 0.5011\pm0.0011 & 0.8111\pm0.0020 & 0.8089\pm0.0017 & 0.8771\pm0.0029 \\
2458923.67628 & 0.1\pm1.1 & 0.5770\pm0.0014 & 0.01119\pm0.00075 & -0.0063\pm0.0034 & 0.50112\pm0.00067 & 0.8102\pm0.0013 & 0.8069\pm0.0012 & 0.8765\pm0.0019 \\
2458982.55918 & 1.0\pm1.5 & 0.5879\pm0.0020 & 0.0012\pm0.0033 & -0.0067\pm0.0014 & 0.5014\pm0.0010 & 0.8055\pm0.0019 & 0.8078\pm0.0017 & 0.8782\pm0.0028 \\
2458989.46752 & -4.0\pm1.2 & 0.5835\pm0.0014 & 0.0040\pm0.0015 & -0.0069\pm0.0012 & 0.50080\pm0.00070 & 0.8021\pm0.0013 & 0.8197\pm0.0012 & 0.8806\pm0.0019 \\
2458999.43482 & -4.3\pm1.3 & 0.5879\pm0.0012 & 0.01061\pm0.00097 & -0.00391\pm0.00065 & 0.50114\pm0.00059 & 0.8036\pm0.0011 & 0.8289\pm0.0010 & 0.8808\pm0.0016 \\
2459006.47245 & -3.2\pm2.0 & 0.5833\pm0.0026 & 0.00551\pm0.00090 & 0.0004\pm0.0010 & 0.5161\pm0.0013 & 0.8045\pm0.0024 & 0.8310\pm0.0022 & 0.8819\pm0.0034 \\
2459012.39093 & 5.1\pm2.0 & 0.5822\pm0.0027 & 0.0052\pm0.0013 & -0.00275\pm0.00088 & 0.4953\pm0.0013 & 0.8008\pm0.0024 & 0.8388\pm0.0022 & 0.8740\pm0.0036 \\
2459024.42111 & 4.5\pm1.8 & 0.5861\pm0.0016 & -0.0125\pm0.0024 & 0.0044\pm0.0011 & 0.49497\pm0.00081 & 0.8032\pm0.0015 & 0.8401\pm0.0014 & 0.8812\pm0.0023 \\
2459036.39359 & 1.3\pm1.3 & 0.5938\pm0.0016 & -0.04560\pm0.00090 & 0.00182\pm0.00076 & 0.49283\pm0.00076 & 0.8057\pm0.0014 & 0.8360\pm0.0013 & 0.8724\pm0.0020 \\
2459053.37775 & {\dots} & 0.5910\pm0.0018 & -0.0569\pm0.0011 & -0.00359\pm0.00093 & 0.48490\pm0.00082 & 0.8038\pm0.0016 & 0.8453\pm0.0015 & 0.8787\pm0.0023 \\
2459053.37782 & {\dots} & 0.5910\pm0.0018 & -0.0569\pm0.0011 & -0.00359\pm0.00093 & 0.48490\pm0.00082 & 0.8038\pm0.0016 & 0.8453\pm0.0015 & 0.8787\pm0.0023 \\
2459061.35479 & 10.6\pm1.5 & 0.5912\pm0.0018 & -0.1054\pm0.0012 & -0.00026\pm0.00084 & 0.48722\pm0.00089 & 0.8002\pm0.0017 & 0.8406\pm0.0016 & 0.8796\pm0.0024 \\
2459214.76769 & -7.0\pm1.5 & 0.5838\pm0.0015 & 0.00374\pm0.00081 & -0.0126\pm0.0042 & 0.50741\pm0.00076 & 0.8095\pm0.0014 & 0.8044\pm0.0013 & 0.8811\pm0.0021 \\
\end{longtable}
}
\end{landscape}

\end{appendix}

\end{document}